\documentstyle[12pt,epsf]{article}
\global\arraycolsep=2pt 
\begin{document}
\def\slash#1{#1 \hskip -0.5em / }
\newcommand{\hqp}{\bar{h}^{(+)}}
\newcommand{\hm}{h^{(-)}}
\newcommand{\hqm}{\bar{h}^{(-)}}
\newcommand{\hp}{h^{(+)}}
\newcommand{\Qqp}{\bar{Q}^{(+)}}
\newcommand{\Qqm}{\bar{Q}^{(-)}}
\newcommand{\Qm}{Q^{(-)}}
\newcommand{\Qp}{Q^{(+)}}
\newcommand{\gaperpmu}{\gamma^{\perp\mu}}
\newcommand{\gaperpnu}{\gamma^{\perp\nu}}
\newcommand{\gaperprh}{\gamma^{\perp\rho}}
\newcommand{\gaperpal}{\gamma^{\perp\alpha}}
\newcommand{\gaperpbe}{\gamma^{\perp\beta}}
\newcommand{\gaperp}{\gamma^{\perp}}
\newcommand{\paperp}{\partial^{\perp}}
\newcommand{\paperpal}{\partial^{\perp\alpha}}
\newcommand{\paperpmu}{\partial^{\perp\mu}}
\newcommand{\paperprh}{\partial^{\perp\rho}}
\newcommand{\Dperp}{i\ D^{\perp}\hskip -1.2em /\;\;\;}
\newcommand{\Dperplr}{i\, \stackrel{\longleftrightarrow}%
{D^{\perp}\hskip -1.2em /\;\;\;}}
\newcommand{\Dperplrrho}{i\, \stackrel{\longleftrightarrow}%
{D^{\perp}_{\rho}}}
\newcommand{\Dperplrsig}{i\, \stackrel{\longleftrightarrow}%
{D^{\perp}_{\sigma}}}
\newcommand{\Drhoperplr}{i\, \stackrel{\longleftrightarrow}%
{D^{\perp\rho}}}
\newcommand{\Dalperplr}{i\, \stackrel{\longleftrightarrow}%
{D^{\perp\alpha}}}
\newcommand{\Dnuperplr}{i\, \stackrel{\longleftrightarrow}%
{D^{\perp\nu}}}
\newcommand{\psidag}{\psi^{\dagger}}
\newcommand{\chidag}{\chi^{\dagger}}
\newcommand{\bfsigma}{\vec{\sigma}}
\newcommand{\bfdif}{i\ \stackrel{\longleftrightarrow}{\vec{D}}}
\newcommand{\bfpa}{i\ \vec{\partial}}
\newcommand{\Pslash}{P\hskip -0.6em /\;}
\newcommand{\pislash}{\pi\hskip -0.5em /\;}
\newcommand{\etaslash}{\eta\hskip -0.5em /\;}
\thispagestyle{empty}
\begin{titlepage}

\begin{flushright}
CERN-TH.7468/94
\end{flushright}

\vspace{0.3cm}

\begin{center}
\Large\bf Heavy Quarkonium Effective Theory
\end{center}

\vspace{0.8cm}

\begin{center}
{\large Thomas Mannel}
and {\large Gerhard A.\ Schuler}\footnote{Heisenberg Fellow.} \\
{\sl Theory Division, CERN, CH-1211 Geneva 23, Switzerland}
\end{center}

\vspace{\fill}

\begin{abstract}
\noindent
We formulate a QCD-based effective theory approach to heavy
quarkonia-like systems as $\bar{c} c$ and $\bar{b} b$ resonances and
$B_c$ states. We apply the method to inclusive decays, working
out a few examples in detail.
\end{abstract}
\vfill
\noindent
CERN-TH.7468/94 \\
September 1994
\end{titlepage}

\section{Introduction}
One of the most important recent developments in the field of
heavy-quark physics has been the formulation of the
static limit for a heavy quark as an effective theory
\cite{SV87}.
The so-called Heavy Quark Effective Theory
(HQET) has put the description of heavy-hadron physics on a
QCD-related and
model-independent basis. Aside from this theoretical progress
there have
been numerous phenomenological applications of this idea, in which the
model dependence has completely disappeared or is substantially
decreased \cite{reviews}.

Most of these applications deal with systems involving a single heavy
quark. In HQET both the particle and
the antiparticle numbers are separately conserved, and the applications
considered so far deal mainly with the one-(anti)particle sector
of HQET. Thus the one-particle sector of HQET is already explored in
some detail, including leading and next-to-leading QCD corrections.

However, there have also been first attempts to deal with the
two-(anti)particle sector and the particle-antiparticle sector
\cite{GK91,KM93}. Here it turns out that complex anomalous
dimensions are obtained by a na\"\i ve analysis.
In HQET the anomalous dimensions in general depend on the
velocities of the heavy quarks and the imaginary parts of the anomalous
dimensions in the two-particle sector diverge as $1/ \sqrt{(vv')^2 - 1}$
for $v \to v'$, if $v$ and $v'$ are the velocities of the two heavy
quarks. Subsequent investigations have shown that this
singularity is related
to the long-range part of the quark-antiquark potential and that one
may remove this divergence by a suitable definition of the multiparticle
states of HQET. In a basis consisting of the redefined states one then
has real anomalous dimensions.

The physical origin of these phases is
the same as that of the well-known Coulomb phases.
The one-gluon (one-photon)
exchange potential decreases too slowly, leading to the well-known
singularities of the form $1/|\vec{v}|$, where $\vec{v}$ is the relative
velocity of the outgoing coloured (charged) particles. From its physical
interpretation as well as from the fact that the divergent phase is a
property of the states, it is clear that these divergent
contributions are
a long-distance effect belonging to the infrared dynamics of the
state with two heavy quarks.

Heavy quarkonium states have to show up as bound states in the
particle-antiparticle sector of HQET. In such a state, however, the two
velocities of the heavy quarks differ by an amount of only
$\Lambda_{QCD} / m$ ($m$ is the heavy-quark mass)
and hence we would rather like to switch to a
description in which the two heavy-quark velocities become equal. This
limit, however, cannot be taken in a na\"\i ve way, since the
static limit of HQET does not reproduce the above phases, which are
related to the potential between the two heavy quarks and hence to the
binding mechanism of the quarkonium state.

It turns out that this limit may be performed if the evolution of the
states is determined not by the static Lagrangian,
but rather by a Lagrangian
consisting of the static HQET part and  the first subleading
spin symmetric contribution to the Lagrangian, i.e. the kinetic
energy operator. In other words, the presence of the divergent
phase forces us to go beyond the static limit, if we want to consider
heavy-quarkonium states, in which the velocities of the heavy quarks
differ only by an amount of order $\Lambda_{QCD} / m$. This in turn
introduces a mass dependence into the lowest-order dynamics such that
the inverse ``Bohr radius'' $\widetilde\Lambda$, which sets a small
non-perturbative scale, is in general not independent of $m$.
For this reason we lose heavy-flavour symmetry, but spin symmetry is
still present, since the kinetic energy operator is still spin
symmetric.

In the present paper we set up an effective theory
approach to processes
involving heavy quarkonia, which is based on the heavy-mass limit of
QCD. We focus on inclusive annihilation-type decays for which we
perform a systematic expansion in powers of $\tilde\Lambda / m$,
up to logarithms, which may be accessed via the renormalization group.

The approach proposed here has many features common to the one
of Bodwin, Braaten and Lepage (BBL)
\cite{BB94}.
Their method is based
on non-relativistic QCD (NRQCD), where the lowest-order dynamics is
determined by the Lagrangian of the Schr\"odinger equation for the
heavy quarks, which is basically that obtained by including
in addition to the static HQET part also the kinetic energy operator.
Based on this they perform an
expansion in terms of $v/c$, where $v$ is the typical relative velocity
of the two heavy quarks. We shall compare the two approaches
between them as well as with previous approaches to inclusive
quarkonium decays (for reviews see e.g.\ \cite{No78,Sc94})
as we go along.

Our general strategy is as follows. We shall first discuss (section 2)
the heavy-mass limit for quarkonia-like systems and show that the
two heavy-particle velocities may be chosen to become equal, once
the subleading kinetic energy operator is added to the
lowest-order dynamics.
This however implies that there will be no mass-independent
static limit for the quarkonia, in other words, the ``binding energy''
$\tilde\Lambda$ will depend on the heavy mass in a
non-trivial way, but it
is still a small scale $\tilde\Lambda \ll m$, so that an expansion in
powers of $\tilde\Lambda / m$ is useful.

The first step to access the inclusive annihilation
decays is to separate long and short distances, where the distance scale
is set by the Compton wavelength of the heavy quark.
Technically speaking,
the first step is to set up an operator product expansion for the
inclusive decay rate of a heavy quarkonium into
light degrees of freedom
(section 3).  In section 4 we write down the
$\tilde\Lambda / m$ expansion for the rate of inclusive
heavy quarkonia annihilation
up to and including $\tilde\Lambda ^2 /m^2$;
up to this order the inclusive decay rate
involves matrix elements of dimension-six and dimension-eight operators.

The coefficients of this expansion
may be calculated in perturbation theory, and some simple examples are
considered in section 5. These perturbatively calculated coefficients
are the ones at the matching scale, i.e. at the scale of the heavy quark
mass. In general the coefficients are scale-dependent,
since the operators
need to be renormalized. In section 6 we consider the renormalization
group flow of the coefficients. Given their values at the matching scale
one may then use the renormalization group to scale down
to some small scale
$\mu$, at which one then tries to estimate the matrix elements of the
corresponding operators. These matrix elements are non-perturbative
quantities, but heavy quark symmetries restrict the
number of independent
parameters. This is studied in section 7.
Finally, we apply this machinery
to some simple examples in section 8. Section 9 contains
a further discussion of previous approaches \cite{No78,Sc94}
as well as our conclusions.
Some technical details are left for the appendices.

\section{Heavy-Mass Limit for Quarkonia}
We shall first set up the heavy-mass limit appropriate for
heavy-quarkonia states. We shall start from the Lagrangian
and the fields of
QCD. We denote the heavy-quark field of full QCD by $Q$ and define
\begin{equation} \label{Qfull1}
Q_v^{(+)} (x) = \exp (im\ vx) Q (x) = h^{(+)}_v (x) + H^{(+)}_v (x)
\ ,
\end{equation}
where $v$ is a velocity ($v^2 = 1$), which is later identified with
the velocity of the heavy hadron. Extracting this phase factor from the
full QCD field $Q$ removes the dominant part $m v $
of the heavy-quark momentum,
since this phase redefinition corresponds to a splitting of the
heavy-quark momentum according to $p = m v + k$, where the residual
momentum $k$ is small, of the order of $\Lambda_{QCD}$.
Furthermore, $h^{(+)}_v$ ($H^{(+)}_v$) is the
large (small) component field, corresponding to the projections
\begin{equation}
h^{(+)}_v = P_+ Q_v^{(+)} , \quad H^{(+)}_v = P_- Q_v^{(+)}
\mbox{ with }
P_\pm = \frac{1}{2} (1\pm \slash{v})
\ .
\end{equation}
The small component field $H^{(+)}_v$ is related
to the large scale $m$;
integrating out $H^{(+)}_v$ from the generating
functional of QCD Green's functions corresponds to the replacement
\begin{equation} \label{Qfull2}
H^{(+)}_v = P_- \left( \frac{1}{2m + ivD} \right) i \slash{D} h^{(+)}_v
\end{equation}
and this yields a non-local Lagrangian of the form \cite{MR92}
\begin{equation} \label{lfull}
{\cal L} = \bar{h}^{(+)}_v (iv D) h^{(+)}_v +
\bar{h}^{(+)}_v i \slash{D} P_- \left( \frac{1}{2m + ivD} \right)
i \slash{D} h^{(+)}_v
\ ,
\end{equation}
which still contains all orders in $1/m$. However, the non-locality
appearing in the second term of (\ref{lfull}) may be expanded into an
infinite series of local terms, which come with increasing powers of
$1/m$. Hence one may in this way establish the desired heavy-mass
expansion. The first few terms of the expansion for the Lagrangian are
\begin{eqnarray}
{\cal L} &=& \bar{h}^{(+)}_v (iv D) h^{(+)}_v +
\left( \frac{1}{2m} \right) \bar{h}^{(+)}_v  i \slash{D}
P_- i \slash{D}  h^{(+)}_v
\nonumber  \\
&& +
\left( \frac{1}{2m} \right)^2 \bar{h}^{(+)}_v i \slash{D} P_-
(-ivD) i \slash{D}  h^{(+)}_v  + \cdots
\ ,
\label{lexp}
\end{eqnarray}
while the field is given by
\begin{equation} \label{>>>}
Q_v^{(+)} = \left( 1 + \frac{1}{2m} P_-  i\slash{D}
               + \frac{1}{4m^2} (-ivD) P_-  i\slash{D}
               + {\cal O} (1/m^3) \right) h^{(+)} (x)
\ .
\end{equation}

The non-local expression (\ref{lfull}) is still equivalent to full QCD;
in particular it is independent of the still arbitrary velocity vector
$v$. In fact, the Lagrangian (\ref{lfull}) is invariant under an
infinitesimal shift of the velocity
\begin{eqnarray}
&& v \to v + \delta v \qquad v \cdot \delta v = 0
\nonumber\\
&& h^{(+)}_v \to h^{(+)}_v + \frac{\delta v \hskip -0.75em /
\hskip 0.25em }{2}
\left(1+ P_- \frac{1}{2m + ivD} i \slash{D} \right) h^{(+)}_v
\nonumber \\
&& iD \to - m \, \delta v \ .
\label{repara}
\end{eqnarray}
This invariance is the so-called reparametrization invariance
\cite{LM92},
which will play some role in what follows.

The Lagrangian (\ref{lfull}) and its expansion in powers of $1/m$
(\ref{lexp}) is the Lagrangian for a heavy quark. In the infinite mass
limit, the quarks and the antiquarks are separated by an infinitely
large mass gap, and hence in order to describe heavy quark-antiquark
systems we have to introduce the antiquark field as a separate field,
the Lagrangian of which is obtained from (\ref{lfull}) or (\ref{lexp})
by the replacement $v \to - v$.

Heavy quarkonia should appear in the  particle-antiparticle
sector of HQET as bound states. The starting point of our considerations
is the Lagrangian of a heavy-quark and a heavy-antiquark field.
In the static limit we obtain from (\ref{lexp})
\begin{equation} \label{lstat}
{\cal L} = \bar{h}_v^{(+)} (ivD) h_v^{(+)} -
           \bar{h}_w^{(-)} (iwD) h_w^{(-)}
\ ,
\end{equation}
where the static quark field $h_v^{(+)}$ ($h_w^{(-)}$)
moves with velocity $v$ ($w$). From (\ref{lstat}) we obtain
the equations of motion
\begin{equation} \label{eom}
(ivD) h^{(+)}_v = 0 \qquad (iwD) h^{(-)}_w = 0
\ .
\end{equation}
However, (\ref{lstat}) or (\ref{eom}) cannot be used to describe heavy
quarkonia states. In order to discuss this we shall consider the matrix
elements
\begin{equation}
\langle A | \bar{Q} (x) \Gamma Q (x) | 0 \rangle
\end{equation}
where $A$ is a state containing a heavy quark and a
heavy antiquark moving with
velocities $v$ and $w$ respectively.
In the static limit this matrix element becomes
\begin{equation} \label{mestat}
\widetilde{\langle A |} \bar{h}_v^{(+)} (x)
\Gamma h_w^{(-)} (x) | 0 \rangle
\end{equation}
where the tilde denotes the static limit of the state.

Matrix elements of this kind have been considered
already in \cite{GK91,KM93}, where the short-distance corrections
have been calculated. It has been
observed that for the na\"\i ve definition of the states,
the anomalous dimension of the current
$\bar{h}_v^{(+)} (x) \Gamma h_w^{(-)} (x)$
acquires an imaginary part of the general structure
\begin{equation} \label{imag}
\mbox{Im }\gamma = f(\alpha_s)
                   \frac{1}{\sqrt{(vw)^2 - 1}}
\end{equation}
as $v \to w$,
where the function $f$ is known as power series in $\alpha_s$ up to
two loops \cite{KM93}
\begin{equation}
f(\alpha_s) = \frac{4}{3} \alpha_s \left(
1 + \frac{\alpha_s}{4 \pi} \left[ \frac{31}{3} -
\frac{10}{9} n_f \right]
+ \cdots \right)
\ .
\end{equation}
The real part of the anomalous dimension vanishes in the limit
$v \to w$ to all orders, since the current
$\bar{h}_v^{(+)} \gamma_\mu h_v^{(+)}$ is a generator of
heavy-flavour symmetry. On the other hand, the anomalous dimensions
of $\bar{h}_v^{(+)} \gamma_\mu h_w^{(+)}$ and
$\bar{h}_v^{(+)} \gamma_\mu h_w^{(-)}$ are related by analytic
continuation and hence the real part is identical.

Thus the only problem is in fact the imaginary part given in
(\ref{imag}), which diverges
in the limit $v \to w$, and this is the major obstacle in
taking the na\"\i ve limit $ v \to w$ in (\ref{lstat}).
Keeping only this imaginary part,
the solution of the renormalization group equation yields a phase
factor
\begin{equation} \label{phase}
\exp (i \phi (vw )) =
\exp\left\{ i \frac{vw}{\sqrt{(vw)^2 - 1}}
\int\limits_{\alpha_s (m)}^{\alpha_s (\mu)} d \alpha
    \frac{f(\alpha)}{\beta(\alpha)} \right\} \quad ; \quad
\beta (\alpha (\mu )) = \mu \frac{\partial}{\partial \mu} \alpha( \mu )
\end{equation}
for matrix elements such as (\ref{mestat}), which is not well
defined in the limit $v \to w$.

It has been pointed out in \cite{KM93} that these phases are related
to the  Coulombic part of the one-gluon exchange and that the phase may
be removed by a suitable definition of the multiparticle states of HQET.
After the redefinition of the states, the anomalous dimensions are real
and well behaved in the limit $v \to w$. Furthermore, the phase in
(\ref{phase}) is related to the ``long-range'' part of the one-gluon
exchange potential and indicates the possibility of having bound states
in some of the quark-(anti)quark channels.

Consequently, the phase appearing in (\ref{phase}) is an infrared
contribution which should be contained in the dynamics of the effective
theory. In other words, if we want to describe heavy quarkonia, the
phase factor is related to the binding mechanism, which is an infrared
effect. In a state such as a quarkonium, the two velocities differ
only by a small amount of order $1/m$ which is a hint that we need to
go beyond the static limit to describe quarkonia states.

In order to see how higher-order terms in the Lagrangian
cure the problem,
we use the reparametrization invariance (\ref{repara}). The expressions
\begin{eqnarray}
&& {\cal V} = v + \frac{\stackrel{\longleftarrow}{iD}}{m}
\mbox{ and }  {\cal W} = w + \frac{iD}{m}
\nonumber\\
&& \tilde{h}_v^{(+)} = \left( 1 + \frac{i \slash{D}}{2 m}
\right) h_v^{(+)}
\mbox{ and }
\tilde{h}_w^{(-)} = \left( 1 + \frac{i \slash{D}}{2 m}
\right) h_w^{(-)}
\end{eqnarray}
are invariant under the reparametrizations $v+\delta v$
and $w+\delta w$,
where this is true for the second line only to order $1/m$.
This observation has been used in \cite{Ne93} to obtain,
for heavy-heavy currents the
renormalization  of the subleading terms from the leading ones. Using
reparametrization invariance, we have
\begin{equation}
\exp (i \phi (vw )) \bar{h}_v^{(+)} \Gamma h_{-w}^{(-)}
\quad \stackrel{\mbox{RPI}}{\longrightarrow} \quad
\tilde{\bar{h}}_v^{(+)}
\exp (i \phi ({\cal V} {\cal W}))
\Gamma \tilde{h}_{-w}^{(-)}
\ .
\end{equation}
If we now consider the limit $v \to w$ we have also
${\cal V} \to {\cal W}$,
which yields in the phase factor the formal expression
\begin{equation}
\exp (i \phi ({\cal V} {\cal W})) \to \exp (i \phi ({\cal V}^2 ))
= \exp\left\{ i \frac{1}{\sqrt{ {\cal V}^2 - 1}}
\int\limits_{\alpha_s (m)}^{\alpha_s (\mu)} d \alpha
    \frac{f(\alpha)}{\beta\alpha} \right\}
\ ,
\end{equation}
where we have only kept the singular term in the last step.

Hence the divergent phase factor may be shifted from the velocities into
the residual momenta, if the equation of
motion of the heavy-quark field is not the one given by (\ref{eom}),
but rather by
\begin{equation} \label{neweom}
\frac{m}{2} \left( {\cal V}^2 - 1 \right) h_v^{(+)} = 0
\quad \mbox{ and } \quad
\frac{m}{2} \left( {\cal W}^2 - 1 \right) h_{-w}^{(-)} = 0
\ .
\end{equation}
In other words, the Lagrangian leading to these equations of motion
is the combination
\begin{eqnarray}
{\cal L}_0 &=& \frac{m}{2}\bar{h}_v^{(+)} ({\cal V}^2 - 1) h^{(+)}_v
+ \frac{m}{2}\bar{h}_{-w}^{(-)} ({\cal W}^2 - 1) h^{(-)}_{-w}
\nonumber\\
           &=& \bar{h}_v^{(+)} (ivD) h^{(+)}_v
             + \bar{h}_v^{(+)} \frac{(i\slash{D})^2}{2m} h^{(+)}_v
             - \bar{h}_w^{(-)} (iwD) h^{(-)}_w
             + \bar{h}_w^{(-)} \frac{(i\slash{D})^2}{2m} h^{(-)}_w
\ ,
\label{lnull}
\end{eqnarray}
where we have replaced $w \to -w$ in the last step.
This expression for the Lagrangian is reparametrization-invariant;
it is also the same invariance which ensures that the
static and the kinetic terms of the
Lagrangian renormalize in the same way. Now one may
perform the limit $v \to w$ without encountering a problem in the
short-distance contributions as in (\ref{imag}). Of course, the
phase (\ref{phase}) has not disappeared: it will show up as a
singularity in the residual relative momentum, which will be generated
by the infrared dynamics as given in (\ref{lnull}). This also
means that it is a long-distance effect and may be absorbed into the
states as discussed in \cite{KM93}.

However, if it is (\ref{lnull}) that determines the leading term of
our expansion,
we will not be able to perform a strict infinite-mass limit, since
now ${\cal L}_0$ depends explicitly on the heavy mass. In the case
of heavy quarkonium, non-perturbative effects will generate binding
of the two heavy objects that will introduce a small scale
$\tilde\Lambda$, which now in general  depends on the heavy mass.

The bottomonium and the charmonium are far from being Coulombic
systems and there is no obvious reason why in the heavy-mass limit
a heavy quarkonium should become Coulombic; still the case of a
Coulombic system is instructive. Neglecting any running of
$\alpha$, the size of such a  Coulombic system is
$R_{Bohr}= 1/(\alpha m)$, which
is large compared to the Compton wavelength $\lambda_Q = 1/m$ and
hence disparate scales
appear allowing for an effective field theory treatment. However, the
small scale $1/R_{Bohr}$ depends on the mass such that it does not
approach a finite limit as $m \to \infty$, even for running $\alpha_s$.

For the heavy quarkonia we shall not assume any type of
potential or binding model, but rather leave
the matrix elements of the operators as non-perturbative entities, which
may not be determined within the effective theory. The only thing that
has to be kept in mind is that these matrix elements depend on the mass,
as can be seen from the  Coulombic example.
The  Coulombic system is, however, an extreme
case, since all scales are set there by the mass of the
constituents; we shall argue below that the mass dependence
might, in reality, be much weaker.

Based on the new equations of motion we get for tree-level
matrix elements
the relations
\begin{equation} \label{releom}
(ivD) h^{(+)}_v = \frac{(iD)^2}{2m} h^{(+)}_v = {\cal O} (1/m) \quad ,
\quad (ivD) h^{(-)}_v = - \frac{(iD)^2}{2m} h^{(-)}_v = {\cal O} (1/m)
\ .
\end{equation}
As we shall see, this means that the static equations of motion
hold up to terms one order higher in the heavy-mass expansion. This
is very similar to what happens in the $v/c$ expansion of
\cite{BB94},
where the right-hand side of (\ref{releom}) is suppressed
by an additional
power in $v/c$.

The  magnetic-moment term of order $1/m$ appearing in the Lagrangian
and all higher-order terms are treated as perturbations
and lead to time-ordered products; these terms may be
identified with the
corrections to the states. The expansion of the fields
$Q_v^{(\pm)}$ leads
to corrections, which are in general local operators.
However, as in any
effective theory we are free to perform field redefinitions
\cite{CC69},
and hence
the expansions (\ref{lexp}) and (\ref{>>>})
are not uniquely defined; one may always move terms appearing in the
expansion of the field into the Lagrangian; these contributions appear
in the Lagrangian as terms that would vanish by a na\"\i ve
use of the equations of motion (\ref{eom}). However, inserted into
time-ordered products they yield local terms according to
\begin{equation} \label{local}
\langle \psi | T \{ \bar{h}^{(+)} \Gamma (ivD) h^{(+)}
                    \bar{h}^{(+)} \Gamma ' h^{(+)} \}
| \psi \rangle  = i \delta^4 (x)
\langle \psi |  \bar{h}^{(+)} \Gamma P_+ \Gamma ' h^{(+)} | \psi \rangle
+ {\cal O} (1/m)
\ ,
\end{equation}
where the higher-order terms appear, since the equation of motion
is not quite the static one. However, this ambiguity of shifting
contributions between the Lagrangian and the fields appears only in
terms of order $1/m^2$ or higher and thus does not affect our arguments
concerning the heavy-mass limit based on reparametrization invariance.

Hence only the combination of (\ref{lexp}) and (\ref{>>>}) has physical
significance and for the present application it is convenient
to use expansions
somewhat different from (\ref{lexp}) and (\ref{>>>}). By a linear
field redefinition we arrive at equivalent expansions, which are up to
$1/m^2$:
\begin{equation}
{\cal L} = {\cal L}_{\mbox{static}} + {\cal L}_I
\end{equation}
where
\begin{eqnarray}
{\cal L}_{\mbox{static}} &=&
    \bar{h}^{(+)}\, ivD\, h^{(-)} -  \bar{h}^{(-)}\, ivD\, h^{(-)}
\nonumber\\
{\cal L}_I &=& \left(\frac{1}{2m} \right) L_1 +
               \left(\frac{1}{2m} \right)^2 L_2
\nonumber \\
           &=& \left(\frac{1}{2m} \right)
                     (K_1  +  G_1 ) +
                     \left(\frac{1}{2m} \right)^2
                     (K_2  + G_2 )  + {\cal O} (1/m^3)
\ .
\label{lexpFW}
\end{eqnarray}
Here we have defined
\begin{eqnarray}
K_1 = K_1^{(+)} + K_1^{(-)} \quad
K_1^{(\pm)} &=&  \bar{h}^{(\pm)} [ (i D)^2 - (ivD)^2 ] h^{(\pm)}
\nonumber \\
G_1 = G_1^{(+)} + G_1^{(-)} \quad
G_1^{(\pm)} &=&  (-i) \bar{h}^{(\pm)} \sigma_{\mu \nu}
               (i D ^\mu)(i D^\nu)  h^{(\pm)}
\nonumber\\
K_2 = K_2^{(+)} + K_2^{(-)} \quad
K_2^{(\pm)} &=& \bar{h}^{(\pm)} [(i D_\mu) ,
[ (-ivD), (iD^\mu) ]] h^{(\pm)}
\nonumber \\
G_2 = G_2^{(+)} + G_2^{(-)} \quad
G_2^{(\pm)} &=&   (-i)  \bar{h}^{(\pm)}
\sigma_{\mu \nu} \{ (i D ^\mu) ,
                               [ (-ivD), (iD^\nu) ] \} h^{(\pm)}
\ .
\end{eqnarray}
The corresponding expansion of the field $\bar{Q}^{(+)}_v$ reads
\begin{eqnarray}
Q_v^{(+)}(x) &=& \left( 1 + \frac{1}{2m} P_-  i\slash{D}
              - \frac{1}{8m^2} (ivD) P_-  i\slash{D} \right.
\nonumber\\
&& \quad \left. - \frac{1}{8m^2} \left( (i D)^2 - (ivD)^2
                       - i \sigma_{\mu \nu} iD^\mu iD^\nu \right)
                       + {\cal O} (1/m^3)
                       \right) h^{(+)} (x)
\ .
\label{fexpFW}
\end{eqnarray}
In fact this is the form that has been obtained from QCD by a sequence
of Foldy-Wouthuysen transformations \cite{KT91}.

The advantage of this form is that now the Lagrangian no longer contains
terms that would vanish by a na\"\i ve use of the equation
of motion; all these terms have been shifted into the expansion of the
field $\bar{Q}^{(+)}_v$. Again the corresponding expressions for the
antiparticle fields are obtained by the replacement $v \to - v$.

In this way one may obtain all terms which explicitly contain a
heavy-quark field.
However, there are more contributions appearing in order
$1/m^2$ which are due to closed loops of heavy quarks in the full theory.
These contributions may be expressed as local
higher-dimensional operators
involving only gluon fields. To order $1/m^2$
there are only two independent operators, leading at tree level to
a contribution ${\cal L}_{glue}$ to the Lagrangian
($g_s^2 = 4\pi\alpha_s$):
\begin{equation}
\left(\frac{1}{2m}\right)^2\; {\cal L}_{glue}
  = \frac{\alpha_s}{30 \pi m^2}
\mbox{ Tr}\{ [ iD_\mu , G^{\mu \nu} ][ iD^\lambda , G_{\lambda \nu} ] \}
+ \frac{i \alpha_s g_s}{360 \pi m^2}
\mbox{ Tr}\{  G^{\mu \nu} [ G_{\nu \rho} ,
G^{\rho}_{\hskip 0.5em \mu} ] \}
\label{Lglue}
\end{equation}
where the gluon field strength is defined as
\begin{equation}
[iD_\mu , iD_\nu ] = i g_s G_{\mu \nu}
\ .
\end{equation}

We may use the equations of motion of the gluon field to
rewrite two of the $1/m^2$
operators as
\begin{eqnarray}
&& \bar{h}^{(+)} [(i D_\mu) , [ (-ivD), (iD^\mu) ]] h^{(+)} =
4 \pi \alpha_s \bar{h}^{(+)} \gamma_\mu T^a  h^{(+)}
        \sum_q \bar{q}  \gamma^\mu T^a  q
\nonumber\\
&& \mbox{ Tr}\{ [ iD_\mu , G^{\mu \nu} ][ iD^\lambda ,
G_{\lambda \nu} ] \}
= - 2 \sum_{qq'} (\bar{q}  \gamma^\mu T^a  q)
                 (\bar{q}' \gamma_\mu T^a  q')
\ .
\end{eqnarray}
In this form they have a simple interpretation: The first one is the
interaction of the heavy quark with the (virtual) light quarks in the
quarkonium, and the second one is the interaction among these (virtual)
light quarks, which is introduced by heavy-quark loops.

In order to calculate short-distance QCD corrections within
this effective
theory, one has to start from the Feynman rules as derived
from ${\cal L}_0$.
The propagator $H(k)$ obtained in this way is
\begin{equation} \label{propeff}
H (k) = P_+ \frac{i}{vk + \frac{1}{2 m} k^2 + i \epsilon}
\end{equation}
and hence contains all orders of $1/m$. However, the reason why we
had to include these higher-order terms was that this removes the
divergent phase occurring in the limit  of small relative velocity.
This phase is a long-distance effect and may be absorbed into the
states, which thus have to evolve according to the dynamics
dictated by ${\cal L}_0$ given in (\ref{lnull}). Once the phase is
removed, we may expand the remaining expression for the short-distance
contribution in powers of $1/m^n$. Hence, as far as practical
calculations are concerned, we may  simply use the static propagator
\begin{equation}
H_{stat} (k) = P_+ \frac{i}{vk + i \epsilon}
\end{equation}
as in usual HQET, which is the leading term of (\ref{propeff}). If we
chose the velocities of the heavy quarks to be equal, then ill-defined
imaginary parts such as
\begin{equation}
\int \frac{d^4 k}{(2 \pi)^4} \frac{1}{k^2} \delta (vk) \delta (v'k)
\to \int \frac{d^4 k}{(2 \pi)^4} \frac{1}{k^2} (\delta (vk))^2
\mbox{ as } v \to v '
\end{equation}
will show up at the one-loop level, which are contributions to the
divergent phase. However, these are absorbed into the states and the
real parts may be calculated simply in the static limit, even for
equal velocities of the heavy quarks.

In this way we may exploit the full machinery of the static limit,
i.e.~of HQET. However, although the short-distance contributions
are calculable in the static limit, the matrix elements of operators
composed of static fields will not be flavour-independent in the case
of a quarkonium, since the states contain a non-trivial mass dependence.

This concludes the set-up for the heavy-mass expansion for
heavy quarkonia.
In the following section we shall apply these ideas to
inclusive annihilation decays of heavy quarkonia.

\section{Separation Between Long and Short Distances}
In this section we shall discuss the general set-up for
a QCD-based calculation of the inclusive annihilation of a
heavy-quarkonium state. The aim is to establish a separation
between long- and short-distance physics; the latter may be
calculated in perturbation theory, while the long-distance
part is non-perturbative and is parametrized in terms of
hadronic matrix elements.

The starting point is the inclusive transition rate $\Gamma$ of a
quarkonium state $|\psi\rangle$, which is given by the optical theorem
in terms of
the forward matrix element of the transition operator $T$
\begin{equation}
\Gamma = 2 \mbox{ Im } \langle \psi | T | \psi \rangle
\ ,
\label{opti}
\end{equation}
which is itself related to the discontinuities across the cuts
of the two-point function of two fields $\Psi$ interpolating the
state $|\psi\rangle$
\begin{equation} \label{green}
G (p) = \int \frac{d^4p}{(2 \pi)^4} e^{-ipx}
\langle 0 | T \{ \Psi (x) \Psi (0) \}  | 0 \rangle
\ .
\end{equation}
For the case at hand, (\ref{green}) has a cut along the real
axis of the complex $p^2$ plane starting at $m_{light}^2$, where
the mass $m_{light}$ corresponds to the lightest hadronic state the
heavy quarkonium may decay into.

On the other hand, the transition rate $\Gamma$ is related to the
discontinuity across the cut at the mass of the heavy quarkonium state,
which is much larger than $m_{light}$. If we consider only decays into
light hadrons we are thus far away from the resonance region, for which
$p^2$ is of order $m_{light}^2$. Using local duality we may calculate
the annihilation of the two heavy quarks into light hadrons
in perturbation theory, since the scale for this process is set by the
heavy-quark mass. In other words, the short-distance piece of the
transition rate $\Gamma$ has an
expansion in inverse powers of the heavy-quark mass, up to logarithms
induced by renormalization, and the coefficients of this expansion
may be calculated in perturbation theory.

The annihilation part of the transition operator $T$ in terms of the
heavy-quark fields is in general given by
\begin{equation}
T = \int d^4 X \, d^4 \rho \, d^4 \xi \, \sum_{C = 1,8} \sum_{j,k}
K_{ij}^{(C)} (X,\rho,\xi) \,
\bar{Q}(X+\rho) \Gamma_j C Q (X - \rho)
    \bar{Q} (-\xi) \bar\Gamma_k C Q (\xi)
\ ,
\end{equation}
where $C$ is a matrix in colour space; $C=1$ ($C=8$)
corresponds to the colour combination
$C \bigotimes C = 1 \bigotimes 1$
($C \bigotimes C = T^a \bigotimes T^a$) and the sum
over $j$ and $k$ runs over the sixteen Dirac matrices. The kernel
$K_{jk}$ depends on the c.m.s.-coordinate $X$ of the heavy-quark pair as
well as on the two relative coordinates $\xi$ and $\rho$.

To set up an expansion in inverse powers of the heavy-quark mass
we perform a phase redefinition of the heavy-quark fields
\begin{equation}
Q (x) = \exp(-im (vx) ) Q_v^{(+)} (x) , \quad
\bar{Q} (x) =  \exp(-im (vx) ) \bar{Q}_{v}^{(-)} (x)
\end{equation}
where the superscript $(+)$ ($(-)$) refers to the
annihilation part of the
fields for quarks (antiquarks). Here $v$ is a velocity vector, which
we chose to be the same for both the quark and the antiquark. Hence we
shall work in the limit we have discussed in the last section and
finally identify the velocity with the velocity of the heavy quarkonium.

This phase redefinition removes the dominant piece of the
space-time dependence
of the heavy-quark field operator; the remaining dependence is only due
to the residual momentum of the heavy quark in the heavy hadron.

Inserting these redefined fields the transition operator takes the form
\begin{eqnarray}
T &=& \int d^4 X \, d^4 \rho \, d^4 \xi \, \sum_{C = 1,8} \sum_{j,k}
K_{ij}^{(C)} (X,\rho,\xi) \, \exp(i2mX v )
\nonumber \\
&& \bar{Q}^{(+)}_v (X+\rho) \Gamma_j C Q^{(-)}_ v  (X - \rho)
    \bar{Q}^{(-)}_v  (-\xi) \bar\Gamma_k C Q^{(+)}_v (\xi)
\ .
\label{tres}
\end{eqnarray}
Introducing the Fourier-transform of the kernel as
\begin{equation}
K_{ij}^{(C)} (X,\rho,\xi) = \int \frac{d^4\tilde{P}}{(2 \pi)^4}
 \frac{d^4 \pi}{(2 \pi)^4}
 \frac{d^4 \eta}{(2 \pi)^4} e^{-i\tilde{P}X} e^{-i\pi \rho} e^{i\eta \xi}
\widetilde{{\cal K}}_{jk}^{(C)} (\tilde{P}, \pi , \eta )
\ ,
\end{equation}
we obtain
\begin{eqnarray}
T &=& \int d^4 X \, d^4 \rho \, d^4 \xi \, \sum_{C = 1,8} \sum_{j,k}
\int \frac{d^4\tilde{P}}{(2 \pi)^4}
 \frac{d^4 \pi}{(2 \pi)^4}
 \frac{d^4 \eta}{(2 \pi)^4}
\widetilde{{\cal K}}_{jk}^{(C)} (\tilde{P}, \pi , \eta )  \,
 e^{-iX(\tilde{P}-2mv)} e^{-i\pi \rho} e^{i\eta \xi}
\nonumber \\
&& \bar{Q}^{(+)}_v (X+\rho) \Gamma_j C Q^{(-)}_ v  (X - \rho)
    \bar{Q}^{(-)}_v  (-\xi) \bar\Gamma_k C Q^{(+)}_v (\xi)
\ .
\end{eqnarray}
The dominant momentum dependence is $2 m v$ and we can remove this
large piece by redefining the c.m.s.\ momentum;
hence we introduce the residual c.m.s.\ momentum
\begin{equation}
P = \tilde{P} - 2mv
\end{equation}
and obtain
\begin{eqnarray}
T &=& \int d^4 X \, d^4 \rho \, d^4 \xi \, \sum_{C = 1,8} \sum_{j,k}
\int \frac{d^4P}{(2 \pi)^4}
 \frac{d^4 \pi}{(2 \pi)^4}
 \frac{d^4 \eta}{(2 \pi)^4}
 {\cal K}_{jk}^{(C)} (m; P , \pi , \eta )  \,
 e^{-iXP } e^{-i\pi \rho} e^{i\eta \xi}
\nonumber \\
&& \bar{Q}^{(+)}_v (X+\rho) \Gamma_j C Q^{(-)}_ v  (X - \rho)
    \bar{Q}^{(-)}_v  (-\xi) \bar\Gamma_k C Q^{(+)}_v (\xi)
\ ,
\end{eqnarray}
where the kernel ${\cal K}_{jk}^{(C)}$ now depends only on small,
i.e.\ residual, momenta and on the heavy mass $m$.

Furthermore, the kernel has an expansion in inverse powers of the
heavy-quark mass $m$, up to logarithms of the form $\ln (m/\mu)$,
where $\mu$ is a factorization scale\footnote{We restrict our
     attention here to the purely perturbative
     aspects of the operator expansion and ignore possible problems
     induced by renormalon poles. These have been discussed recently
     in the context of the heavy mass expansion in \cite{renormalons}.}.
The transition rate $\Gamma$
is independent of the factorization scale, and thus the scale dependence
of the expansion of the kernel has to be compensated by the corresponding
dependence of the matrix elements, which appear in the expansion.
As usual, this scale dependence is governed by the renormalization group,
and hence one may recover the logarithmic dependence of the transition
rate on the heavy-quark mass $m$.

The expansion of the kernel thus takes the general form
($\lambda = \mu / m$)
\begin{eqnarray}
{\cal K}_{jk}^{(C)} (m; P , \pi , \eta ) &=& R_{jk}^{(C)} (\lambda ) +
\frac{1}{2m} \left( P_\mu S_{ij}^{(C) \mu} (\lambda ) +
                    \pi_\mu T_{ij}^{(C) \mu} (\lambda )+
                    \eta_\mu U_{ij}^{(C) \mu} (\lambda ) \right)
\nonumber \\
&+& \left(\frac{1}{2m}\right)^2
    \left( P_\mu P_\nu S_{ij}^{(C)(1)\mu \nu }(\lambda ) +
           P_\mu \pi_\nu S_{ij}^{(C)(2) \mu \nu } (\lambda )+
           P_\mu \eta_\nu S_{ij}^{(C)(3) \mu \nu } (\lambda )\right.
\nonumber \\
&& \hphantom{+ \left(\frac{1}{2m}\right)^2}
   \vphantom{\left(\frac{1}{2m}\right)^2}
        + \pi_\mu P_\nu T_{ij}^{(C)(1) \mu \nu } (\lambda )+
           \pi_\mu \pi_\nu T_{ij}^{(C)(2) \mu \nu } (\lambda )+
           \pi_\mu \eta_\nu T_{ij}^{(C)(3) \mu \nu } (\lambda )
\nonumber \\
&& \hphantom{+ \left(\frac{1}{2m}\right)^2}
   \vphantom{\left(\frac{1}{2m}\right)^2}
 \left. + \eta_\mu P_\nu U_{ij}^{(C)(1) \mu \nu } (\lambda)+
           \eta_\mu \pi_\nu U_{ij}^{(C)(2) \mu \nu } (\lambda )+
          \eta_\mu \eta_\nu U_{ij}^{(C)(3) \mu \nu }(\lambda ) \right)
\nonumber \\
&+& {\cal O} (1/m^3)
\ .
 \label{kernel}
\end{eqnarray}
The powers of the momenta will simply yield matrix elements of
operators involving derivatives. For later use it is convenient to
introduce those that correspond to the residual-c.m.s.-momentum
(RCM)
\begin{equation} \label{RCM}
i \partial_\mu
\left(\bar{Q}^{(+)}_v \Gamma Q^{(-)}_v \right)
=  \left(\bar{Q}^{(+)}_v
                           \Gamma (iD_\mu) Q^{(-)}_v \right)
+  \left(\bar{Q}^{(+)}_v (i \stackrel{\longleftarrow}{D}_\mu )
                           \Gamma Q^{(-)}_v \right)
\end{equation}
and to the residual-relative-momentum  (RRM)
\begin{equation} \label{RRM}
\left(\bar{Q}^{(+)}_v (i \stackrel{\longleftrightarrow}{D}_\mu )
                           \Gamma Q^{(-)}_v \right)
= \left(\bar{Q}^{(+)}_v \Gamma (iD_\mu) Q^{(-)}_v \right)
  - \left(\bar{Q}^{(+)}_v (i \stackrel{\longleftarrow}{D}_\mu )
                           \Gamma Q^{(-)}_v \right)
\end{equation}
where $D = \partial - ig {\cal A}$ is the QCD covariant derivative for
fields in the fundamental representation.
For the other bilinear $\bar{Q}^{(-)}_v \cdots Q^{(+)}_v$ we define
the derivatives in the same way.

Inserting (\ref{kernel}) into the expression for the $T$ operator
and using the definitions (\ref{RCM}) and (\ref{RRM}) one
finds for the transition rate $\Gamma$ up to and including order $1/m^2$:
\begin{eqnarray}
\langle \psi | T | \psi \rangle &=& \sum_{C = 1,8} \sum_{j,k}
\left\{ \vphantom{\frac{1}{1}} R_{jk}^{(C)} (\mu/m )
\langle \psi | [\bar{Q}^{(+)}_v  \Gamma_j C Q^{(-)}_ v ]
[\bar{Q}^{(-)}_v  \bar\Gamma_k C Q^{(+)}_v] | \psi \rangle_\mu
\right.
\nonumber \\
&+& \frac{1}{2m} \left[
S_{ij}^{(C) \mu} (\lambda )
\langle \psi | [ i \partial_\mu (\bar{Q}^{(+)}_v \Gamma_j C Q^{(-)}_ v)]
[\bar{Q}^{(-)}_v  \bar\Gamma_k C Q^{(+)}_v ]| \psi \rangle_\mu \right.
\nonumber \\
&&
\hphantom{+ \frac{1}{2m}} +
T_{ij}^{(C) \mu} (\lambda  )
\langle \psi | [(\bar{Q}^{(+)}_v \Gamma_j
(i \stackrel{\longleftrightarrow}{D}_\mu )
C Q^{(-)}_ v]
[\bar{Q}^{(-)}_v  \bar\Gamma_k C Q^{(+)}_v] | \psi \rangle_\mu
\nonumber \\
&& \left.
\hphantom{+ \frac{1}{2m}}  +
U_{ij}^{(C) \mu} (\lambda  )
\langle \psi | [(\bar{Q}^{(+)}_v \Gamma_j C Q^{(-)}_ v)]
[\bar{Q}^{(-)}_v  \bar\Gamma_k C
(i \stackrel{\longleftrightarrow}{D}_\mu) Q^{(+)}_v] | \psi \rangle_\mu
\right]
\nonumber \\
&+& \left(\frac{1}{2m}\right)^2 \left[
S_{ij}^{(C)(1) \mu \nu} (\lambda  )
\langle \psi | [ i \partial_\mu (\bar{Q}^{(+)}_v \Gamma_j C Q^{(-)}_ v)]
[ i \partial_\mu (\bar{Q}^{(-)}_v
\bar\Gamma_k C Q^{(+)}_v)] | \psi \rangle_\mu \right.
\nonumber \\
&& \hphantom{+ \left(\frac{1}{2m}\right)^2} +
S_{ij}^{(C)(2) \mu \nu} (\lambda  )
\langle \psi | [ i \partial_\mu (\bar{Q}^{(+)}_v \Gamma_j
(i \stackrel{\longleftrightarrow}{D}_\nu )
C Q^{(-)}_ v)]
[\bar{Q}^{(-)}_v
\bar\Gamma_k C Q^{(+)}_v] | \psi \rangle_\mu
\nonumber \\
&& \hphantom{+ \left(\frac{1}{2m}\right)^2} +
S_{ij}^{(C)(3) \mu \nu} (\lambda  )
\langle \psi | [ i \partial_\mu (\bar{Q}^{(+)}_v \Gamma_j
C Q^{(-)}_ v)]
[\bar{Q}^{(-)}_v
\bar\Gamma_k C (i \stackrel{\longleftrightarrow}{D}_\nu )
Q^{(+)}_v] | \psi \rangle_\mu
\nonumber \\
&& \hphantom{+ \left(\frac{1}{2m}\right)^2} +
T_{ij}^{(C)(1) \mu \nu} (\lambda  )
\langle \psi | [  \bar{Q}^{(+)}_v \Gamma_j
(i \stackrel{\longleftrightarrow}{D}_\mu )
C Q^{(-)}_ v)]
[ i \partial_\nu (\bar{Q}^{(-)}_v
\bar\Gamma_k C Q^{(+)}_v ] | \psi \rangle_\mu
\nonumber \\
&& \hphantom{+ \left(\frac{1}{2m}\right)^2} +
T_{ij}^{(C)(2) \mu \nu} (\lambda  )
\langle \psi | [  \bar{Q}^{(+)}_v \Gamma_j
(i \stackrel{\longleftrightarrow}{D}_\mu)
(i \stackrel{\longleftrightarrow}{D}_\nu)
C Q^{(-)}_ v ]
[\bar{Q}^{(-)}_v
\bar\Gamma_k C Q^{(+)}_v] | \psi \rangle_\mu
\nonumber \\
&& \hphantom{+ \left(\frac{1}{2m}\right)^2} +
T_{ij}^{(C)(3) \mu \nu} (\lambda  )
\langle \psi | [ \bar{Q}^{(+)}_v \Gamma_j
(i \stackrel{\longleftrightarrow}{D}_\mu )
C Q^{(-)}_ v]
[\bar{Q}^{(-)}_v
\bar\Gamma_k C (i \stackrel{\longleftrightarrow}{D}_\nu )
Q^{(+)}_v] | \psi \rangle_\mu
\nonumber \\
&& \hphantom{+ \left(\frac{1}{2m}\right)^2} +
U_{ij}^{(C)(1) \mu \nu} (\lambda  )
\langle \psi | [ i \partial_\nu ( \bar{Q}^{(+)}_v \Gamma_j
C Q^{(-)}_ v)]
[ \bar{Q}^{(-)}_v
\bar\Gamma_k C
(i \stackrel{\longleftrightarrow}{D}_\mu )
Q^{(+)}_v ] | \psi \rangle_\mu
\nonumber \\
&& \hphantom{+ \left(\frac{1}{2m}\right)^2} +
U_{ij}^{(C)(2) \mu \nu} (\lambda  )
\langle \psi | [  \bar{Q}^{(+)}_v \Gamma_j
(i \stackrel{\longleftrightarrow}{D}_\nu)
C Q^{(-)}_ v ]
[\bar{Q}^{(-)}_v
\bar\Gamma_k C
(i \stackrel{\longleftrightarrow}{D}_\mu )
Q^{(+)}_v] | \psi \rangle_\mu
\nonumber \\
&& \hphantom{+ \left(\frac{1}{2m}\right)^2} +
\left. \vphantom{\frac{1}{1} }
U_{ij}^{(C)(3) \mu \nu} (\lambda  )
\langle \psi | [ \bar{Q}^{(+)}_v \Gamma_j
C Q^{(-)}_ v]
[\bar{Q}^{(-)}_v
(i \stackrel{\longleftrightarrow}{D}_\mu )
(i \stackrel{\longleftrightarrow}{D}_\nu )
Q^{(+)}_v] | \psi \rangle_\mu \right]
\nonumber \\
&+& \left. \vphantom{\frac{1}{1} } {\cal O} (1/m^3) \right\}
 \label{kern}
\end{eqnarray}
where the subscript $\mu$ at the matrix elements indicates their
renormalization point and
all field operators have to be taken at $x=0$. In what follows we
shall not display the dependence on $x$ any more, if $x=0$.
Furthermore, we have replaced the ordinary
derivatives appearing in the Taylor expansion
of the matrix elements by covariant ones
in order to ensure gauge invariance; this could be implemented from the
very beginning by defining the $T$ operator including appropriate
Wilson-line operators.

The lengthy expression (\ref{kern}) is the
most general short-distance expansion
for the matrix elements of the $T$ operator for the case of heavy
quarkonia decay up to $1/m^2$, and this is the
first ingredient for a $1/m$ expansion of the transition rate $\Gamma$.
In a second step, one has to expand the matrix elements appearing
in (\ref{kern}) in powers of
$1/m$.

\section{$1/m$ Expansion for Inclusive Quarkonia Annihilation}
In this section we write down the complete heavy-mass expansion
for the matrix elements appearing in (\ref{kern}). The result we
are aiming at is an expansion of the transition rate
$\Gamma$ of the form
\begin{equation} \label{gamexp}
\Gamma = \Gamma_0 + \left(\frac{1}{2m}\right) \Gamma_1 +
                    \left(\frac{1}{2m}\right)^2 \Gamma_2 + \cdots
\end{equation}
where we shall explicitly construct the terms up to and including
$1/m^2$. In this expansion, the ratio of two successive $\Gamma_j$ is
set by a small scale
\begin{equation}
\frac{\Gamma_{j+1}}{\Gamma_j} = \widetilde{\Lambda}
\end{equation}
where  $\widetilde{\Lambda}$ is small compared to the heavy-quark
mass. However, there is some
difference between the heavy quarkonia and
the heavy-light systems. In the latter the parameter
corresponding to $\widetilde{\Lambda}$ is
$\bar{\Lambda} = M_{hadron} - m$, which becomes
independent of the heavy quark mass in the heavy-mass limit.

As we have discussed above, in heavy quarkonia systems this is not
true, since in order to perform the  limit $v \to w$ we had to include
subleading terms of the $1/m$ expansion into ${\cal L}_0$, which
determines the dynamics of the states. In this way a non-trivial
dependence on $m$ is introduced into the scale
$\widetilde{\Lambda}$ induced by the
binding of the two heavy quarks. Nevertheless, this is a small scale
and the expansion will be useful.

Since this means that we are not able to perform a static limit for
a heavy quarkonium, we shall no longer have heavy-flavour symmetry.
However, if we define the scale  $\widetilde{\Lambda}$ in a similar
way as $\bar{\Lambda}$ for heavy light systems
\begin{equation}
\widetilde{\Lambda} =  \frac
{iv\partial  \langle 0 |
  \bar{h}_v^{(-)} \Gamma h_v^{(+)} | \psi (v) \rangle }
{ \langle 0 | \bar{h}_v^{(-)} \Gamma h_v^{(+)} |
  \psi (v) \rangle }
\end{equation}
we may get some idea of the flavour dependence by looking at the
level spacings of the $(n{}^3 S_1)$ quarkonia of the $c$ and the $b$
quarks. Here it is a well-known fact that the level spacings are
almost the same in the $(\bar{c} c)$ and $(\bar{b} b)$ system, which
we take as a hint that $\widetilde{\Lambda}$ is only weakly dependent
on the heavy flavour. We shall return to this point when we discuss
heavy-quark symmetries in section 7.

The way we shall set up the expansion (\ref{gamexp}) is to use the
expansions  (\ref{lexpFW}) and  (\ref{fexpFW}). The fields
$Q_v^{(\pm)}$ are expanded as in (\ref{fexpFW}) while the chromomagnetic
term $G_1$ and all terms of order $1/m^2$ of the Lagrangian are treated
as perturbations, which lead to time-ordered products.
Up to and including
$1/m^2$ only the double insertion of $G_1$ and the single insertion of
the order $1/m^2$ terms of the Lagrangian will play a role.

The leading-order term $\Gamma_0$ in the heavy-mass expansion
is given by forward-matrix-elements of
dimension-six operators involving
the static fields $h^{(\pm)}$. In general,
there are four dimension-six operators, two with
colour structure $1 \bigotimes 1$
\begin{eqnarray}
A_1^{(1)} &=& \left(\bar{h}^{(+)} \gamma_5 h^{(-)} \right)
              \left(\bar{h}^{(-)} \gamma_5 h^{(+)} \right)
\nonumber\\
A_2^{(1)} &=& \left(\bar{h}^{(+)} \gamma_\mu h^{(-)} \right)
              \left(\bar{h}^{(-)} \gamma^\mu h^{(+)} \right)
\label{As}
\end{eqnarray}
and the corresponding two operators
$A_1^{(8)}$, $A_2^{(8)}$
with colour structure $T^a \bigotimes T^a$. In fact,
due to the projections
$P_\pm = (1\pm\slash{v})/2$ implicit in the static operators these are
the only spinor structures that can emerge (see appendix for details).

Thus the leading term $\Gamma_0$ is always given as a linear combination
of the four dimension-six operators $A_i^{(C)}$
(the factor $1/2$ arises from (\ref{opti}))
\begin{equation} \label{Leading}
\frac{1}{2}\,
\Gamma_0  =   \sum_{i=1}^2 \sum_{C = 1,8} {\cal C} (A_i^{(C)})
\langle \psi | A_i^{(C)} | \psi \rangle
\ ,
\end{equation}
where the coefficients ${\cal C} (A_i^{(C)})$ depend on the specific
process, i.e.\ the kernel $K_{ij}$ appearing in (\ref{kern}). We shall
give a few examples of this ``matching procedure'' below.

In higher orders of the heavy-mass expansion matrix elements of
higher-dimensional operators appear from the
expansion of the $Q_v^{(\pm)}$,
which have increasing powers of covariant derivatives acting on the
static fields and it is convenient to use the RCM and RRM derivatives
introduced in (\ref{RCM}) and (\ref{RRM}).

All dimension-seven operators either vanish due to symmetries or
they are proportional to $(ivD)$. The latter may be rewritten
using the equations of motion; in the static case their tree-level
matrix element vanishes, while one obtains a dimension-eight
contribution, if we include the higher-order terms of the equation of
motion (\ref{eom}). Whatever choice one prefers, there are no tree-level
matrix elements of local dimension-seven  operators.
The only contribution
which is allowed at order $1/m$ is a single insertion
of the first-order chromomagnetic-moment operator
\begin{equation}
\langle \psi | T\left\{ [G_1^{(+)}(x)+G_1^{(-)}(x)[
               A_i^{(C)} \right\} | \psi \rangle
\end{equation}
for which we shall show below that it vanishes due to spin symmetry.
Thus for the inclusive heavy quarkonia decays there is
no contribution of
order $\widetilde{\Lambda}/m$.

The second-order term in the $1/m$ expansion has contributions from both
the Lagrangian and the local operators. The local
dimension-eight operators
may be classified according to the derivatives defined above.
The first set of operators consists of the contributions with two
RCM derivatives (RCM $\times$ RCM). There are in general six independent
contributions of this type, three of which are colour $1 \bigotimes 1$
and the other four are the corresponding $T^a \bigotimes T^a$
operators. The colour $1 \bigotimes 1$ operators are
\begin{eqnarray}
B_1^{(1)} &=& \left[ i \partial_\mu
              \left(\bar{h}^{(+)} \gamma_5 h^{(-)} \right) \right]
              \left[ i \partial^\mu
              \left(\bar{h}^{(-)} \gamma_5 h^{(+)} \right) \right]
\nonumber\\
B_2^{(1)} &=& \left[ i \partial_\mu
              \left(\bar{h}^{(+)} \gamma^\mu h^{(-)} \right) \right]
              \left[ i \partial^\nu
              \left(\bar{h}^{(-)} \gamma_\nu h^{(+)} \right) \right]
\nonumber\\
B_3^{(1)} &=& \left[ i \partial_\mu
              \left(\bar{h}^{(+)} \gamma^\nu h^{(-)} \right) \right]
              \left[ i \partial^\mu
              \left(\bar{h}^{(-)} \gamma_\nu h^{(+)} \right) \right]
\ .
\label{Bidef}
\end{eqnarray}
Note that one may flip the RCM derivatives from one heavy-quark bilinear
to the other, thereby picking up a total derivative, which will not
contribute to the forward matrix elements we shall consider. This is the
reason why no terms appear in which both derivatives act on
the same heavy-quark bilinear; these may be rewritten into
the $B$ operators.

In the same way one obtains six colour $1 \bigotimes 1$ RCM $\times$
RRM operators
\begin{eqnarray}
C_1^{(1)} &=& \left\{
              \left(\bar{h}^{(+)} \gamma_5
              (i \stackrel{\longleftrightarrow}{D}_\mu )
               h^{(-)} \right)
              \left[ i \partial^\mu
              \left(\bar{h}^{(-)} \gamma_5 h^{(+)} \right) \right]
              + \mbox{h.c.} \right\}
\nonumber\\
C_2^{(1)} &=& \left\{
              \left(\bar{h}^{(+)}
              (i \stackrel{\longleftrightarrow}{\slash{D}} )
               h^{(-)} \right)
              \left[ i \partial^\mu
              \left(\bar{h}^{(-)} \gamma_\mu h^{(+)} \right) \right]
              + \mbox{h.c.} \right\}
\nonumber\\
C_3^{(1)} &=& \left\{
              \left(\bar{h}^{(+)} \gamma_\nu
              (i \stackrel{\longleftrightarrow}{D}_\mu )
               h^{(-)} \right)
              \left[ i \partial^\mu
              \left(\bar{h}^{(-)} \gamma_\nu h^{(+)} \right) \right]
              + \mbox{h.c.} \right\}
\nonumber\\
C_4^{(1)} &=& \left\{
              \left(\bar{h}^{(+)} \gamma_\nu
              (i \stackrel{\longleftrightarrow}{D}_\mu )
               h^{(-)} \right)
              \left[ i \partial^\nu
              \left(\bar{h}^{(-)} \gamma_\mu h^{(+)} \right) \right]
              + \mbox{h.c.} \right\}
\nonumber\\
C_5^{(1)} &=& \left\{i \varepsilon^{\alpha \beta \kappa \mu}
                     v_\alpha
              \left(\bar{h}^{(+)} \gamma_\beta
              (i \stackrel{\longleftrightarrow}{D}_\kappa )
               h^{(-)} \right)
              \left[ i \partial_\mu
              \left(\bar{h}^{(-)} \gamma_5 h^{(+)} \right) \right]
              + \mbox{h.c.} \right\}
\nonumber\\
C_6^{(1)} &=& \left\{i \varepsilon^{\alpha \beta \kappa \mu}
                     v_\alpha
              \left(\bar{h}^{(+)} \gamma_5
              (i \stackrel{\longleftrightarrow}{D}_\kappa )
               h^{(-)} \right)
              \left[ i \partial_\mu
              \left(\bar{h}^{(-)} \gamma_\beta h^{(+)} \right) \right]
              + \mbox{h.c.} \right\}
\label{Cidef}
\end{eqnarray}
and the corresponding $T^a \bigotimes T^a$ operators.

Finally, there are  22 RRM $\times$ RRM operators, half of which
are $1 \bigotimes 1$ and the other half are the corresponding colour
$T^a \bigotimes T^a$. These operators fall into two categories, namely
(for the colour singlets)
\begin{eqnarray}
D_1^{(1)} &=&
              \left(\bar{h}^{(+)} \gamma_5
              (i \stackrel{\longleftrightarrow}{D}_\mu )
               h^{(-)} \right)
              \left(\bar{h}^{(-)} \gamma_5
              (i \stackrel{\longleftrightarrow}{D}^\mu )
               h^{(+)} \right)
\nonumber\\
D_2^{(1)} &=&
              \left(\bar{h}^{(+)}
              (i \stackrel{\longleftrightarrow}{\slash{D}} )
               h^{(-)} \right)
              \left(\bar{h}^{(-)}
              (i \stackrel{\longleftrightarrow}{\slash{D}})
               h^{(+)} \right)
\nonumber\\
D_3^{(1)} &=&
              \left(\bar{h}^{(+)} \gamma_\mu
              (i \stackrel{\longleftrightarrow}{D}_\nu )
               h^{(-)} \right)
              \left(\bar{h}^{(-)} \gamma^\mu
              (i \stackrel{\longleftrightarrow}{D}^\nu )
               h^{(+)} \right)
\nonumber\\
D_4^{(1)} &=&
              \left(\bar{h}^{(+)} \gamma_\nu
              (i \stackrel{\longleftrightarrow}{D}_\mu )
               h^{(-)} \right)
              \left(\bar{h}^{(-)} \gamma^\mu
              (i \stackrel{\longleftrightarrow}{D}^\nu )
               h^{(+)} \right)
\nonumber\\
D_5^{(1)} &=& \left\{ i \varepsilon^{\alpha \beta \kappa \mu}
                     v_\alpha
              \left(\bar{h}^{(+)} \gamma_\beta
              (i \stackrel{\longleftrightarrow}{D}_\kappa )
               h^{(-)} \right)
              \left(\bar{h}^{(-)} \gamma_5
              (i \stackrel{\longleftrightarrow}{D}_\mu )
               h^{(+)} \right)  + \mbox{h.c.} \right\}
\label{Didef}
\end{eqnarray}
and
\begin{eqnarray}
E_1^{(1)} &=& \left\{
              \left(\bar{h}^{(+)} \gamma_5
              (i \stackrel{\longleftrightarrow}{D})^2
               h^{(-)} \right)
              \left(\bar{h}^{(-)} \gamma_5
               h^{(+)} \right)
              + \mbox{h.c.} \right\}
\nonumber\\
E_2^{(1)} &=& \left\{
              \left(\bar{h}^{(+)}
              (i \stackrel{\longleftrightarrow}{\slash{D}})
              (i \stackrel{\longleftrightarrow}{D}_\mu)
               h^{(-)} \right)
              \left(\bar{h}^{(-)} \gamma^\mu
               h^{(+)} \right)
              + \mbox{h.c.} \right\}
\nonumber\\
E_3^{(1)} &=& \left\{
              \left(\bar{h}^{(+)}
              (i \stackrel{\longleftrightarrow}{D}_\mu)
              (i \stackrel{\longleftrightarrow}{\slash{D}})
               h^{(-)} \right)
              \left(\bar{h}^{(-)} \gamma^\mu
               h^{(+)} \right)
              + \mbox{h.c.} \right\}
\nonumber\\
E_4^{(1)} &=& \left\{
              \left(\bar{h}^{(+)}
              (i \stackrel{\longleftrightarrow}{D} )^2
              \gamma_\mu
               h^{(-)} \right)
              \left(\bar{h}^{(-)} \gamma^\mu
               h^{(+)} \right)
              + \mbox{h.c.} \right\}
\nonumber\\
E_5^{(1)} &=& \left\{ i \varepsilon^{\alpha \beta \kappa \mu}
                     v_\alpha
              \left(\bar{h}^{(+)} \gamma_\beta
              (i \stackrel{\longleftrightarrow}{D}_\kappa )
              (i \stackrel{\longleftrightarrow}{D}_\mu )
               h^{(-)} \right)
              \left(\bar{h}^{(-)} \gamma_5
               h^{(+)} \right)
              + \mbox{h.c.} \right\}
\nonumber\\
E_6^{(1)} &=& \left\{ i \varepsilon^{\alpha \beta \kappa \mu}
                     v_\alpha
              \left(\bar{h}^{(+)} \gamma_5
              (i \stackrel{\longleftrightarrow}{D}_\kappa )
              (i \stackrel{\longleftrightarrow}{D}_\mu )
               h^{(-)} \right)
              \left(\bar{h}^{(-)} \gamma_\beta
               h^{(+)} \right)
              + \mbox{h.c.} \right\}
\ .
\label{Eidef}
\end{eqnarray}
Thus there are in general 40 dimension-eight operators.
Note that we have
already dropped the operators which vanish because
of the equations of motion for the heavy quark.

The general expression for $\Gamma_2$ is thus given by a linear
combination of the 40 local dimension-eight operators
and the non-local
contributions from the time-ordered products with the Lagrangian
(recall the factor $1/(4m^2)$ in the definitions
(\ref{gamexp}), (\ref{lexpFW}), and (\ref{Lglue})
and the factor $1/2$ in (\ref{opti}))
\begin{eqnarray}
\frac{1}{2}\,
\Gamma_2 &=& \sum_{i=1}^3 \sum_{C = 1,8} {\cal C} (B_i^{(C)})
             \langle \psi | B_i^{(C)} | \psi \rangle
           + \sum_{i=1}^6 \sum_{C = 1,8} {\cal C} (C_i^{(C)})
             \langle \psi | C_i^{(C)} | \psi \rangle
\nonumber \\
         &+& \sum_{i=1}^5 \sum_{C = 1,8} {\cal C} (D_i^{(C)})
             \langle \psi | D_i^{(C)} | \psi \rangle
          +  \sum_{i=1}^6 \sum_{C = 1,8} {\cal C} (E_i^{(C)})
             \langle \psi | E_i^{(C)} | \psi \rangle
\nonumber \\
         &+&
          \sum_{i=1}^2 \sum_{C = 1,8}
          {\cal C}  (A_i^{(C)}) (-i) \int d^4 x \,
\langle \psi | T\left\{ L_2 (x) A_i^{(C)} (0) \right\} | \psi \rangle
\nonumber \\
 &+&  \frac{(-i)^2}{2}
          \sum_{i=1}^2 \sum_{C = 1,8}
         {\cal C} (A_i^{(C)})
          \int d^4 x \, d^4 y \,
\langle \psi | T\left\{ G_1 (x) G_1 (y) A_i^{(C)} (0)
        \right\} | \psi \rangle
\nonumber \\
         &+&
          \sum_{i=1}^2 \sum_{C = 1,8}
          {\cal C}  (A_i^{(C)}) (-i) \int d^4 x \,
\langle \psi | T\left\{
               {\cal L}_{glue} (x) A_i^{(C)} (0) \right\} | \psi \rangle
 \label{second}
\end{eqnarray}
where the coefficients in the time-ordered product terms are again
given by the lowest-order coefficients ${\cal C} (A_i^{(C)})$. This
expression is still quite general and simplifies once the heavy-quark
symmetry and the fact that we have to take forward matrix elements
is taken into account.

\section{The Operator Coefficients to Leading Non-Trivial Order}
\begin{figure}
   \vspace{0.5cm}
   \epsfysize=4cm
   \centerline{\epsffile{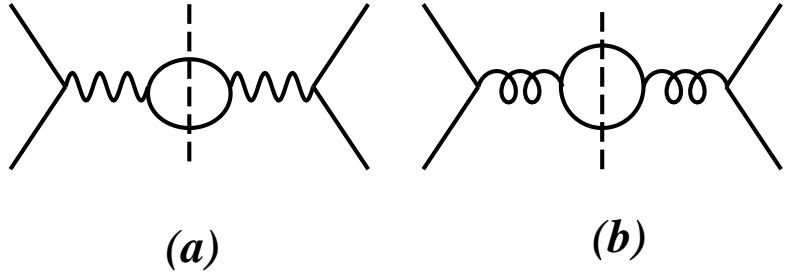}}
   \centerline{\parbox{16cm}{\caption{\label{fig3}
Fermion loop diagrams. Diagram (a) contributes to the electromagnetic
kernel for the quarkonia decays into $e^+ e^-$, diagram (b) contributes
to the strong interaction kernel for the quarkonia decays into light
hadrons. The dashed line means that the intermediate states are on-shell.
}}}
\end{figure}

In this section we shall calculate two simple kernels in order to
show how our method is applied, namely the fermion
loop diagrams depicted in fig.~\ref{fig3}.

These diagrams yield contributions of order $\alpha$ (fig.~\ref{fig3}a)
and $\alpha_s(m)$ (fig.~\ref{fig3}b). Calculating the electromagnetic
contribution (fig.~\ref{fig3}a) one obtains for the kernel
\begin{equation}
{\cal K}_{\mu \nu} (P,\pi,\eta) =
         - \frac{ 4 \pi}{3 P^2} \alpha^2 c_Q^2 g_{\mu \nu}
\ ,
\end{equation}
where $c_Q$ is the charge of the heavy quark in units of
the electron charge.
Furthermore, $P = p + \bar{p}$ is the sum of the momenta of
the quark $p$ and
the antiquark $\bar{p}$.
Hence only the Dirac matrix combination $\gamma_\mu
\otimes \gamma^\mu$
contributes in the sum over the Dirac matrices in (\ref{kern}).
The kernel
is expanded using
$$
p = v \sqrt{m^2 - p_\perp^2} + p_\perp = v
\left( m   -   \frac{1}{2m} p_\perp^2 + \cdots \right) + p_\perp
$$
and the result is matched to the operators by identifying the momenta
with the derivatives appearing in the operators (see appendix for more
details). The only non-vanishing coefficients are
\begin{eqnarray}
&& {\cal C}^{ee} (A_2^{(1)},m ) =  {\cal C}^{ee} (B_3^{(1)},m ) =
    - \frac{\pi}{3m^2} \alpha^2 c_Q^2
\nonumber \\
&& {\cal C}^{ee} (E_2^{(1)},m ) = {\cal C}^{ee} (E_3^{(1)},m ) =
- \frac{\pi}{12m^2} \alpha^2 c_Q^2
\nonumber \\
&& {\cal C}^{ee} (E_4^{(1)},m ) =
- \frac{\pi}{6m^2} \alpha^2 c_Q^2 .
\end{eqnarray}
{}From this one may obtain the coefficients for the quark-loop diagram
of fig.~\ref{fig3}b by the replacements
$c_Q^2 \alpha^2 \to n_f \alpha_s^2 (m) /2$ in the coefficients
and $1 \bigotimes 1 \to T^a \bigotimes T^a$ in the operators,
where $n_f$ is the number
of light flavours which are allowed in the quark loop. In this way we
obtain
\begin{eqnarray}
&& {\cal C}^{qq} (A_2^{(8)},m ) =  {\cal C}^{qq} (B_3^{(8)},m ) =
    - \frac{\pi}{6m^2} \alpha_s^2 (m) n_f
\nonumber \\
&& {\cal C}^{qq} (E_2^{(8)},m ) = {\cal C}^{qq} (E_3^{(8)},m ) =
- \frac{\pi}{24m^2} \alpha_s^2 (m) n_f
\nonumber \\
&& {\cal C}^{qq} (E_4^{(8)},m ) =
- \frac{\pi}{12m^2} \alpha_s^2 (m) n_f
\ .
\end{eqnarray}
The matching calculation yields these coefficients
at the scale $\mu = m$ and hence it is $\alpha_s$ taken at this scale
that enters the expression. We have indicated the $\mu$ dependence
of the coefficients by an additional
argument for these functions. A change of this scale is governed by the
renormalization group of the effective theory, which is discussed in the
next section.
%

\section{QCD Evolution of the Coefficients {\cal C}}
In general QCD corrections render the operators and the coefficients
scale-dependent in such a way that the transition rate $\Gamma$ is
scale-independent. Schematically this may be written as
\begin{equation}
\Gamma = \sum_i {\cal C}_i (\mu) \langle {\cal O}_i  \rangle|_\mu
\end{equation}
where $\langle \rangle|_\mu$ means that
the matrix element is normalized at scale
$\mu$. The scale dependence of the matrix elements and the coefficients
${\cal C}_i$ is governed by the anomalous-dimension matrix $\gamma$
\begin{equation}
\mu \frac{\partial}{\partial \mu} \langle {\cal O}_i \rangle|_\mu
= - \gamma_{ij} \, \langle {\cal O}_j \rangle|_\mu
\ ,
\end{equation}
which is obtained in the standard fashion from the ultraviolet
divergences of the matrix elements. In order to have the transition
rate scale-independent, the coefficients
$ {\cal C}_i (\mu) $ have to obey the renormalization group equation
\begin{equation}
\mu \frac{d}{d \mu} {\cal C}_j = {\cal C}_i \, \gamma_{ij}
\ ,
\label{evol}
\end{equation}
where the initial condition at the scale $\mu = m$ is given
from the matching calculation, i.e.\ the
calculation of the hard kernels as performed in the last
section.

The four dimension-six operators $A_i^{(C)}$ have a vanishing
anomalous-dimension matrix and hence there is no mixing
between these four operators.
Keeping the velocity of the quark operator different from the one for
the antiquark we find that the anomalous-dimension matrix -- after the
redefinition of the states -- vanishes as
\begin{equation}
\gamma^{(6)}_{ij} \propto \omega r(\omega) - 1
\ ,
\end{equation}
where $\omega$ is the product of the two velocities and
\begin{equation}
r(x) = \frac{1}{\sqrt{x^2 - 1}} \ln (x + \sqrt{x^2 - 1})
\ .
\end{equation}
Hence the coefficient functions ${\cal C} (A_i^{(C)})$ are
scale-independent.

In a similar way we find that there is no mixing between the local
dimension-eight operators. In order to access the renormalization of
the non-local time-ordered product terms, we have to first study the
renormalization of the terms in the Lagrangian. The renormalization of
the first-order Lagrangian is \cite{FGL92}
\begin{eqnarray}
\left. \vphantom{\int} K_1 \right|_\mu = C_0 C_1
\left. \vphantom{\int} K_1 \right|_m & \qquad&
        C_1 = 1
\nonumber\\
\left. \vphantom{\int} G_1 \right|_\mu = C_0 C_2
\left. \vphantom{\int} G_1 \right|_m & \qquad&
        C_2 = \eta^{-9/(33-2n_f)}
\end{eqnarray}
where
\begin{equation}
\eta = \frac{\alpha_s (\mu)}{\alpha_s (m)}
\end{equation}
and $C_0$ is the wave function renormalization constant of the static
field, which reads in Feynman gauge
\begin{equation}
C_0 = \eta ^{8/(33-2n_f)}
\ .
\end{equation}
The subscripts $\mu$ and $m$ label at which scale the matrix elements,
in which the operators are inserted, have to be normalized. We note that
$C_1=1$ ensures that both terms of the Lagrangian
(\ref{lnull}) renormalize
in the same way; in other words, the wave-function renormalization is
the same in both the static and the effective theory based on
(\ref{lnull}).

Of the second-order Lagrangian we need only the contributions
$K_2^{(\pm)}$ and ${\cal L}_{glue}$, since there will be
no contribution from
a single insertion of $G_2^{(\pm)}$ due to spin symmetry. Furthermore,
we shall not include the renormalization of the second-order
Lagrangian,
since its contribution is hard to estimate, no matter at which scale
we consider the matrix elements.

\begin{figure}
   \epsfysize=4cm
   \centerline{\epsffile{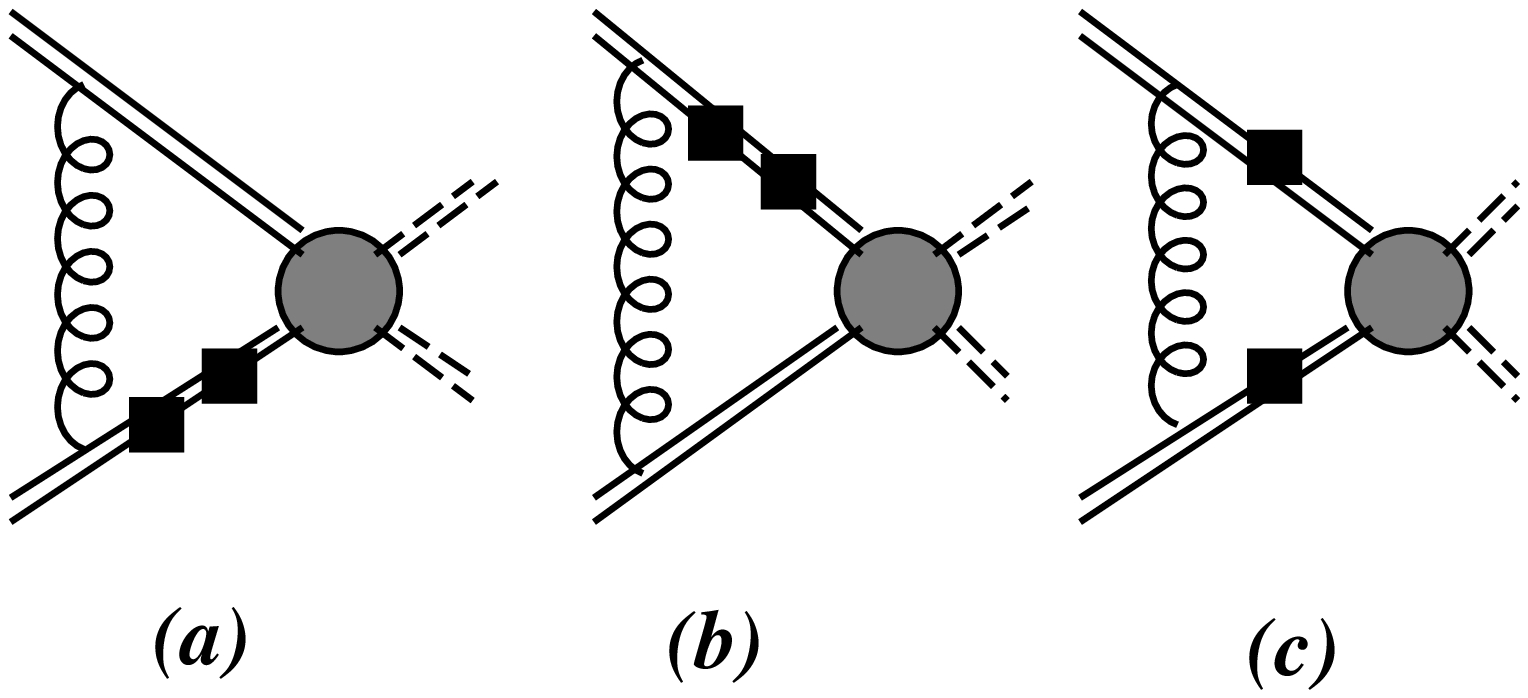}}
   \epsfysize=4cm
   \centerline{\epsffile{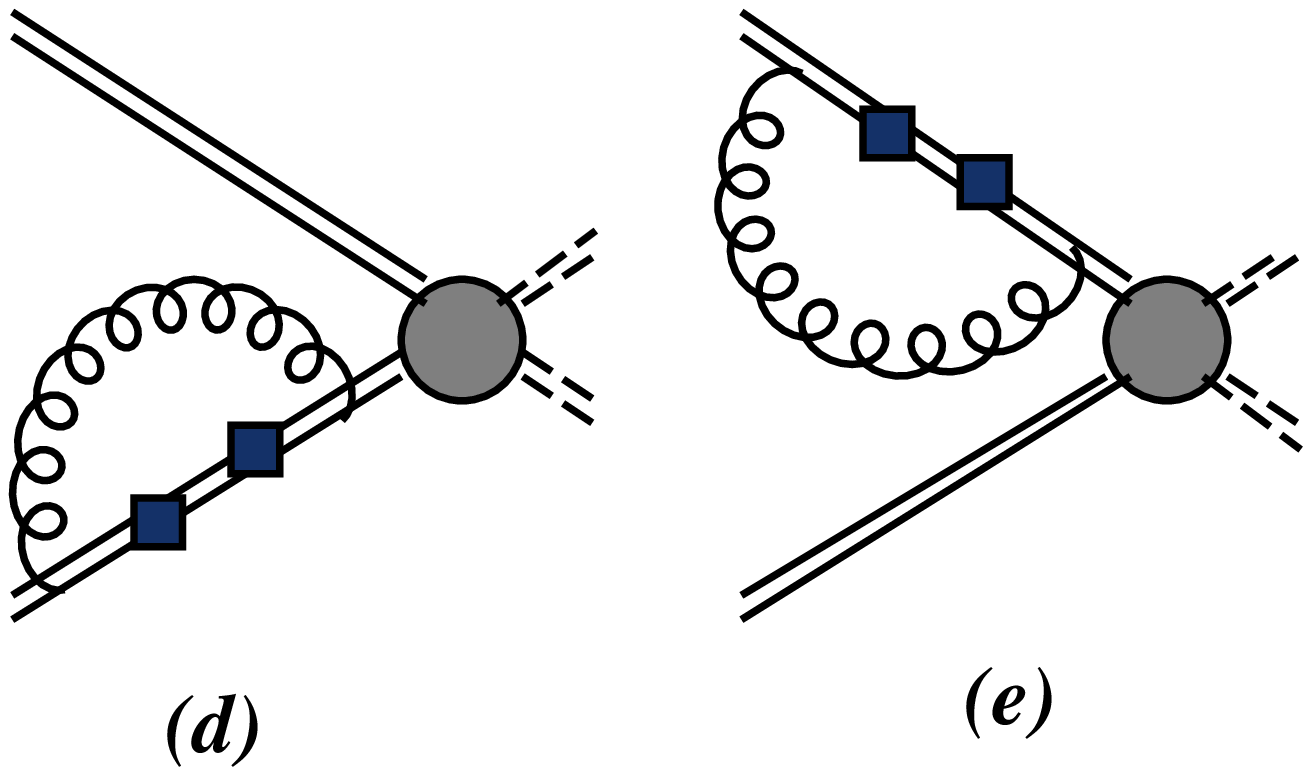}}
   \centerline{\parbox{16cm}{\caption{\label{fig1}
Feynman diagrams leading to mixing between the non-local and the local
operators. The square boxes correspond to insertions of the first-order
kinetic energy operator $K_1$ and the shaded blob corresponds to
the dimension-six operators $A_i^{(C)}$. The diagrams (a) to (e) have to
be calculated for all permutations of the external lines.
}}}
\end{figure}

One important effect of the renormalization in the present application
is the mixing of the time-ordered products of the
Lagrangian into the local operators, since non-local operators may
require a local counterterm. The only contribution that leads to
an ultraviolet divergence is a double insertion of the first-order
kinetic energy operator $K_1$. We define the non-local operators
\begin{equation}
T_i^{(C)} = \frac{(-i)^2}{2}
    \int d^4 x \, d^4 y \,  T\left\{ K_1 (x) K_1 (y)
A_i^{(C)} (0) \right\}
\ ,
\end{equation}
and the one-loop diagram types leading to
ultraviolet-divergent contributions
are depicted in fig.~\ref{fig1}. Diagrams $(a)$ to
$(c)$ are one-particle
irreducible pieces, while $(d)$ and $(e)$ may be
interpreted as the mixing
of the time-ordered product $T\{ K_1 (x) K_1 (0) \}$ into
local terms of the
Lagrangian.

We define two sets of seven operators, the first corresponds to the spin
singlet, the second to the spin triplet coupling
\begin{eqnarray}
({\cal O}^{[1]}_i) & = & (T_1^{(1)}, T_1^{(8)}, D_1^{(1)},
E_1^{(1)}, B_1^{(8)},
                D_1^{(8)}, E_1^{(8)} )
\nonumber\\
({\cal O}^{[3]}_i) & = & (T_2^{(1)}, T_2^{(8)}, D_3^{(1)},
E_4^{(1)}, B_3^{(8)},
                D_3^{(8)}, E_4^{(8)} )
\ .
\end{eqnarray}
These two sets have the same anomalous-dimension matrix
and we find
\begin{equation}
\gamma_{ij} = - \frac{g^2}{6 \pi^2} \left[
\begin{array}{c c c c c c c}
0   &   0      &   0   &  -2  &  0   &  4  &   0    \\
0   &   0      &  8/9  &   0  & -6   & 5/3 & -7/6   \\
0   &   0      &   0   &   0  &  0   &  0  &   0    \\
0   &   0      &   0   &   0  &  0   &  0  &   0    \\
0   &   0      &   0   &   0  &  0   &  0  &   0    \\
0   &   0      &   0   &   0  &  0   &  0  &   0    \\
\quad 0 \quad  & \quad 0 \quad & \quad 0 \quad  & \quad 0 \quad
&  \quad 0 \quad   &  \quad 0 \quad  &   \quad 0 \quad
\end{array} \right]
\ .
\end{equation}
Using the anomalous-dimension matrix, one may calculate the
scale dependence
of the coefficient functions
\begin{eqnarray}
({\cal C}_i^{[1]}) & = & ({\cal C}(T_1^{(1)}), {\cal C}(T_1^{(8)}),
                {\cal C}(D_1^{(1)}), {\cal C}(E_1^{(1)}),
                {\cal C}(B_1^{(8)}),
                {\cal C}(D_1^{(8)}), {\cal C}(E_1^{(8)}))
\nonumber\\
({\cal C}_i^{[3]}) & = & ({\cal C}(T_2^{(1)}), {\cal C}(T_2^{(8)}),
                {\cal C}(D_3^{(1)}), {\cal C}(E_4^{(1)}),
                {\cal C}(B_3^{(8)}),
                {\cal C}(D_3^{(8)}), {\cal C}(E_4^{(8)}))
\ ,
\end{eqnarray}
by solving the renormalization group equation (\ref{evol})
with the initial conditions given at $\mu_0 = m$.
The solutions may be expressed in terms of the value of the
coefficients at the scale $m$ and one obtains
\begin{eqnarray}
{\cal C}(T_1^{(1)}, \mu) &=& {\cal C}(T_1^{(1)}, m)
= {\cal C}(A_1^{(1)},m)
\nonumber \\
{\cal C}(T_1^{(8)}, \mu) &=& {\cal C}(T_1^{(8)}, m)
= {\cal C}(A_1^{(8)},m)
\nonumber \\
{\cal C}(D_1^{(1)}, \mu) &=& {\cal C}(D_1^{(1)}, m)
+ \frac{32}{9} \frac{1}{33 - 2n_f} {\cal C}(T_1^{(8)}, m)
\ln \eta
\nonumber \\
{\cal C}(E_1^{(1)}, \mu) &=& {\cal C}(E_1^{(1)}, m)
-  \frac{8}{33 - 2n_f} {\cal C}(T_1^{(1)}, m)
\ln \eta
\nonumber \\
{\cal C}(B_1^{(8)}, \mu) &=& {\cal C}(B_1^{(8)}, m)
- \frac{24}{33 - 2n_f} {\cal C}(T_1^{(8)}, m)
\ln \eta
\nonumber \\
{\cal C}(D_1^{(8)}, \mu) &=& {\cal C}(D_1^{(8)}, m)
+  \frac{16}{33 - 2n_f} {\cal C}(T_1^{(1)}, m)
\ln \eta
\nonumber \\
&& \qquad \qquad
+  \frac{20}{3} \frac{1}{33 - 2n_f} {\cal C}(T_1^{(8)}, m)
\ln \eta
\nonumber \\
{\cal C}(E_1^{(8)}, \mu) &=& {\cal C}(E_1^{(8)}, m)
- \frac{14}{3} \frac{1}{33 - 2n_f} {\cal C}(T_1^{(8)}, m)
\ln \eta
\end{eqnarray}
and the same solution for the second set of coefficients.

The non-trivial evolution of all the coefficients of the local
operators
is driven by the coefficient of the time-ordered product corresponding
to a double insertion of $K_1$; in turn, this coefficient is simply
the one of the dimension-six operator with the corresponding spin
structure.

The scale $\mu_0$ is the matching scale, which we chose to
be $\mu_0 = m$.
On the other hand, we want to study the matrix elements of the operators
at some lower scale, say the scale of the binding energy of the
quarkonium,
where we expect to have a reasonable approximation using a wave-function
model. The evolution equation (\ref{evol}) allows us to change the
renormalization scale, thereby inducing mixing of the non-local
operators
into the local ones.

\section{Forward Matrix Elements and Heavy Quark Symmetry}
Once we have scaled down to the small scale of the order of the
binding energy of the quarkonium, we have
to evaluate the forward matrix elements appearing in
(\ref{second}). This requires in general non-perturbative input, which
has to be supplied by other methods such as lattice gauge theory or
by model estimates.
However, we may use heavy-quark symmetry to count the number
of independent non-perturbative parameters.

The Lagrangian ${\cal L}_0$ given in (\ref{lnull}) still has
the heavy-quark spin symmetry.
As a consequence, the heavy quarkonia systems fall in general into
spin symmetry quartets: For a given orbital angular momentum $\ell$ and
radial excitation quantum number $n$, the four states
(in the spectroscopic notation ${}^{2S+1}\ell_J$)
\begin{equation}
[n{}^1 \ell_\ell \quad n{}^3 \ell_{\ell-1} \quad n{}^3 \ell_\ell \quad
 n{}^3 \ell_{\ell+1} ]
\label{fourstates}
\end{equation}
form such a spin symmetry quartet. An exception are the $S$ waves
($\ell = 0$), for which the three polarization directions of the
$n{}^3S_1$ and the $n{}^1S_0$ form the spin symmetry quartet.

In order to exploit the consequences of the spin symmetry for
the inclusive hadronic decays
we shall use the trace formalism. We denote with $|Y_\ell\rangle$
the spin symmetry quartet consisting of the spin singlet and the
spin triplet for a given orbital angular momentum $\ell$. The
coupling of the heavy-quark spins may be represented by the
matrices
\begin{equation}
H_Y (v)= \left\{ \begin{array}{l}
      P_+ \gamma_5 \mbox{ for the spin singlet} \\
      P_+ \slash{\epsilon} \mbox{ for the spin triplet}
              \end{array} \right.
\ ,
\end{equation}
where $P_+ = (1+\slash{v})/2$ is the projection of
the ``large components''.
Note that the matrices $H_Y$ are independent of $\ell$.

For the dimension-six operators this implies
\begin{equation}
\langle Y_\ell | \left(\bar{h}^{(+)}\, C\,
                     \Gamma\, h^{(-)} \right)
              \left(\bar{h}^{(-)} \, C\, \Gamma ' \,
               h^{(+)} \right) | Y_\ell \rangle =
a_\ell^{(C)}  \mbox{ Tr}\left\{ \overline{H}_Y \Gamma \right\}
                  \mbox{ Tr}\left\{ \Gamma ' H_Y  \right\}
\ ,
\label{dimsix}
\end{equation}
which means that there are only two independent parameters for a given
$\ell$ describing the
dimension-six matrix elements, namely $a_\ell^{(1)}$ and $a_\ell^{(8)}$.
In the study of exclusive non-leptonic decays of heavy-light mesons,
matrix elements of four-quark operators are usually estimated in
vacuum insertion. Applying this approximation
to (\ref{dimsix}) suggests that $a_0^{(1)}$ is the dominant
coefficient.

The generic dimension-eight colour $1 \bigotimes 1$ operators
may be written in terms of these representations as
\begin{eqnarray}
\langle Y_\ell | [ i \partial^\mu \left(\bar{h}^{(+)}
                     \Gamma h^{(-)} \right)][
                   i \partial^\nu
              \left(\bar{h}^{(-)} \Gamma '
               h^{(+)} \right)] | Y_\ell \rangle & = &
b^{\mu \nu}_\ell  \mbox{ Tr}\left\{ \overline{H}_Y \Gamma \right\}
                  \mbox{ Tr}\left\{ \Gamma ' H_Y  \right\}
\nonumber \\
\left[ \langle Y_\ell |  \left(\bar{h}^{(+)}
              (i \stackrel{\longleftrightarrow}{D}^\mu )
                     \Gamma h^{(-)} \right)[
                   i \partial^\nu
              \left(\bar{h}^{(-)} \Gamma '
               h^{(+)} \right)] | Y_\ell \rangle
+ \mbox{ h.c.} \right] & = &
c^{\mu \nu}_\ell  \mbox{ Tr}\left\{ \overline{H}_Y \Gamma \right\}
                  \mbox{ Tr}\left\{ \Gamma ' H_Y  \right\}
\nonumber\\
\langle Y_\ell |  \left(\bar{h}^{(+)}
              (i \stackrel{\longleftrightarrow}{D}^\mu )
                     \Gamma h^{(-)} \right)
              \left(\bar{h}^{(-)} \Gamma '
              (i \stackrel{\longleftrightarrow}{D}^\nu )
               h^{(+)} \right)] | Y_\ell \rangle & = &
d^{\mu \nu}_\ell  \mbox{ Tr}\left\{ \overline{H}_Y \Gamma \right\}
                  \mbox{ Tr}\left\{ \Gamma ' H_Y  \right\}
\nonumber\\
\left[ \langle Y_\ell |  \left(\bar{h}^{(+)}
              (i \stackrel{\longleftrightarrow}{D}^\mu )
              (i \stackrel{\longleftrightarrow}{D}^\nu )
                     \Gamma h^{(-)} \right)
              \left(\bar{h}^{(-)} \Gamma '
               h^{(+)} \right)] | Y_\ell \rangle
+ \mbox{ h.c.} \right] & = &
e^{\mu \nu}_\ell  \mbox{ Tr}\left\{ \overline{H}_Y \Gamma \right\}
                  \mbox{ Tr}\left\{ \Gamma ' H_Y  \right\}
\ .
\nonumber\\
\label{dimeight}
\end{eqnarray}
The tensors $b^{\mu \nu}_\ell$, $c^{\mu \nu}_\ell$,
$d^{\mu \nu}_\ell$ and $e^{\mu \nu}_\ell$ have to be
constructed from the
velocity vector $v$ and the metric tensor. Furthermore, if one
contracts one of the indices with $v$, one may use the equation of
motion of the fields $h_v^{(\pm)}$; in the static limit this simply
vanishes, while with an equation of motion as (\ref{neweom})
one obtains a
term one order higher in the $1/m$ expansion, which may be dropped,
since
we work only to order $1/m^2$. Consequently, the tensors are given by
\begin{eqnarray}
 b^{\mu \nu}_\ell = b_\ell (g^{\mu \nu} - v^\mu v^\nu) \quad & , \quad &
 c^{\mu \nu}_\ell = c_\ell (g^{\mu \nu} - v^\mu v^\nu)\ ,
\nonumber\\
 d^{\mu \nu}_\ell = d_\ell (g^{\mu \nu} - v^\mu v^\nu) \quad & , \quad &
 e^{\mu \nu}_\ell = e_\ell (g^{\mu \nu} - v^\mu v^\nu) \ .
\end{eqnarray}
Spin symmetry thus places a very strong restriction on the local matrix
elements, since for a given orbital angular momentum $\ell$
there are only eight
non-perturbative parameters $b_\ell^{(1)}, \cdots e_\ell^{(1)}$ and the
corresponding octet partners $b_\ell^{(8)}, \cdots e_\ell^{(8)}$, which
describe all the forward matrix elements
of the local dimension-eight operators given above.

{}From an estimate based on vacuum insertion one would guess that
$b_0^{(1)}$, $d_1^{(1)}$ and $e_0^{(1)}$ dominate all other coefficients;
in particular all the $c_\ell^{(C)}$ are expected to be small, if
vacuum insertion makes any sense.

In a very similar way one may analyse the spin symmetry structure of the
non-local terms. The kinetic energy terms are spin symmetric; furthermore
there is no contribution from insertions of $K_1$, since this is already
contained in the dynamics of the states; however, we note that this does
not mean that $K_1$ does not show up at all: A double insertion of $K_1$
mixes under renormalization into local dimension-eight operators,
but this is
a short-distance effect which was calculated in the last section.

For an insertion of the second-order kinetic energy term $K_2$ we obtain
from the trace formalism
\begin{equation}
(-i) \int d^4 x \, \langle Y_\ell |  T \{ K_2 (x)
                  \left(\bar{h}^{(+)}  \Gamma h^{(-)} \right)
                  \left(\bar{h}^{(-)} \Gamma '  h^{(+)} \right)
                  | Y_\ell \rangle =
k_2^{(1)}  \mbox{ Tr}\left\{ \overline{H}_Y \Gamma \right\}
                  \mbox{ Tr}\left\{ \Gamma ' H_Y  \right\}
\end{equation}
and a corresponding expression for the colour
$T^a \bigotimes T^a$ contribution.

Furthermore, spin symmetry implies that a single insertion of a
chromomagnetic moment operator vanishes. We have ($j = 1,2$):
\begin{eqnarray}
&& (-i) \int d^4 x \, \langle Y_\ell |  T \{ G_j (x)
                  \left(\bar{h}^{(+)}  \Gamma h^{(-)} \right)
                  \left(\bar{h}^{(-)} \Gamma '  h^{(+)} \right)
                  | Y_\ell \rangle =
\nonumber\\
&& g_j^{\mu \nu}  \left[
\mbox{ Tr }\left\{ \overline{H}_Y \sigma_{\mu \nu} P_+ \Gamma \right\}
                  \mbox{ Tr}\left\{ \Gamma ' H_Y  \right\} +
\mbox{ Tr }\left\{ \overline{H}_Y  \Gamma P_- \sigma_{\mu \nu} \right\}
                  \mbox{ Tr}\left\{ \Gamma ' H_Y  \right\} \right.
\nonumber\\
&& \left. \quad\quad +
\mbox{ Tr}\left\{ \overline{H}_Y  \Gamma \right\}
\mbox{ Tr}\left\{ \Gamma ' P_+ \sigma_{\mu \nu} H_Y  \right\} +
\mbox{ Tr}\left\{ \overline{H}_Y  \Gamma \right\}
\mbox{ Tr}\left\{ \sigma_{\mu \nu} P_- \Gamma ' H_Y  \right\} \right]
\ .
\end{eqnarray}
Thus the tensor $g_j^{\mu \nu}$ has to be antisymmetric, but it may
only be built from the velocity vector and the metric tensor; from these
only symmetric combinations are possible, and hence these contributions
have to vanish.

Finally, there is also a double insertion of the chromomagnetic
moment operator
$G_1$. Analysing the spin structure of this term gives
\begin{eqnarray}
\frac{(-i)^2}{2} \int d^4 x \, d^4 y \,
                  \langle Y_\ell |  T \{ G_1 (x) G_1 (y)
                  \left(\bar{h}^{(+)}  \Gamma h^{(-)} \right)
                  \left(\bar{h}^{(-)} \Gamma '  h^{(+)} \right)
                  | Y_\ell \rangle =  S^{\mu \nu \alpha \beta}
\nonumber\\
 \left[
\mbox{ Tr }\left\{ \overline{H}_Y \sigma_{\mu \nu} P_+
                   \sigma_{\alpha \beta} P_+ \Gamma \right\}
                  \mbox{ Tr }\left\{ \Gamma ' H_Y  \right\} +
\mbox{ Tr }\left\{ \overline{H}_Y P_+ \sigma_{\mu \nu} \Gamma
                   P_- \sigma_{\alpha \beta}   \right\}
                  \mbox{ Tr }\left\{ \Gamma ' H_Y  \right\}
 + \cdots \right]
\end{eqnarray}
where the ellipses stand for all other possible insertions of the
two sigma-matrices. The tensor $S$ now has to be antisymmetric and it
may only be built from the metric tensor $g_{\mu \nu}$, since we have
$P_\pm \sigma_{\mu \nu} v^\mu P_\pm = 0$ as well  as
$P_\pm \sigma_{\mu \nu} v^\mu P_\mp = 0$. Hence we have only a single
parameter for this contribution
\begin{eqnarray}
\frac{(-i)^2}{2} \int d^4 x \, d^4 y \,
                  \langle Y_\ell |  T \{ G_1 (x) G_1 (y)
                  \left(\bar{h}^{(+)}  \Gamma h^{(-)} \right)
                  \left(\bar{h}^{(-)} \Gamma '  h^{(+)} \right)
                  | Y_\ell \rangle =  G^{(1)}
\nonumber\\
  \left[
\mbox{ Tr }\left\{ \overline{H}_Y \sigma_{\mu \nu} P_+
                   \sigma^{\mu \nu} P_+ \Gamma \right\}
                  \mbox{ Tr }\left\{ \Gamma ' H_Y  \right\} +
\mbox{ Tr }\left\{ \overline{H}_Y P_+ \sigma_{\mu \nu} \Gamma
                   P_- \sigma^{\mu \nu}   \right\}
                  \mbox{ Tr }\left\{ \Gamma ' H_Y  \right\}
 + \cdots \right]
\end{eqnarray}
and a corresponding relation for the colour
combination $T^a \bigotimes T^a$.

The constraints from spin symmetry thus allow us to drop
the operators $C_5^{(C)}$, $C_6^{(C)}$, $D_5^{(C)}$, $E_5^{(C)}$ and
$E_6^{(C)}$ in a calculation for the total rate.
Of course, the contribution of ${\cal L}_{glue}$, (\ref{Lglue}),
is spin-symmetric, since it involves only the light degrees
of freedom.
Furthermore, the only
contribution of chromomagnetic-moment
operators is the double insertion of $G_1$, which in turn is the only
spin-symmetry-violating contribution that appears.

At the end of this section we want to discuss the flavour symmetry.
Let us recall that for heavy-light systems both the spin and
the flavour symmetries hold. As
already mentioned several times, the flavour symmetry is broken
in the present case of heavy quarkonia, because
the states acquired a mass dependence.
The interesting question arises about
the amount of flavour-symmetry violation. As a simple example,
we study the simpler problem of flavour-symmetry breaking for
mass splittings.
In the heavy-mass limit the four states in
(\ref{fourstates}) should be degenerate;
splitting between the members of the spin-symmetry quartet is
induced by spin-orbit and spin-spin interactions. We may analyse this
by looking at the mass of a heavy quarkonium in terms of the
Lagrangian
\begin{eqnarray}
M (n{}^{2S+1} \ell _J ) &=& 2 m + \widetilde\Lambda (n,\ell) +
\frac{1}{2 m} \langle \psi | [G_1^{(+)} + G_1^{(-)}] | \psi \rangle
\nonumber \\
&&+  \left( \frac{1}{2 m} \right)^2
\langle \psi | [K_2^{(+)} + K_2^{(-)}+ G_2^{(+)} +
G_2^{(-)}] | \psi \rangle
\nonumber \\
&&+  (-i) \left( \frac{1}{2 m} \right)^2 \int d^4 x \,
\langle \psi | T \{ [G_1^{(+)}(x) + G_1^{(-)}(x)]
                    [G_1^{(+)} + G_1^{(-)}] \} | \psi \rangle
\nonumber\\ &&
+ {\cal O} (1/m^3)
\label{masssplit}
\ .
\end{eqnarray}
Note that no insertions of $K_1^{(\pm)}$ appear, since this term is
already included in ${\cal L}_0$.

The first contribution $\widetilde\Lambda (n,\ell)$ is the ``binding
energy'', as one would obtain from the solution of the
Schr\"odinger-type of equations corresponding to ${\cal L}_0$.
This parameter
can only depend on the quantum numbers
$n$ and $\ell$, since spin symmetry is still unbroken.

Breaking of spin symmetry occurs first at order $1/m$; it is given by
the expectation value of the first-order
chromomagnetic moment operator.
In a non-relativistic language, this would correspond to a coupling
of the form $(\vec{s}^{\,(+)} + \vec{s}^{\,(-)}) \cdot \vec{B}$,
where $\vec{B}$
is the chromomagnetic field, which is created by the orbital motion.
Hence we expect $\vec{B} \propto \vec{L}$, where $\vec{L}$ is
the orbital
angular momentum. In other words, the first-order term is a spin-orbit
coupling term.

The spin symmetric local terms of order $1/m^2$ are corrections to
$\widetilde\Lambda (n,\ell)$, while the second-order chromomagnetic
operator yields a correction to the spin-orbit term. The time-ordered
product of the two first-order chromomagnetic moment operators will
give (aside from a correction term to  $\widetilde\Lambda (n,\ell)$)
a spin-spin coupling term, which in a non-relativistic form is
$\vec{s}^{(+)} \cdot \vec{s}^{(-)}$ for $\ell = 0$.
Hence we arrive at a mass formula
of the form (for $\ell = 0$):
\begin{eqnarray}
M (n{}^{2S+1} \ell _J ) &=& 2 m + \widetilde\Lambda (n,\ell) +
\frac{\omega (n,\ell)}{2m}\,  \frac{1}{2}\,
[J (J+1) - \ell (\ell + 1) - S ( S +1 )]
\nonumber\\
& & \quad +
\frac{ \tau(n, \ell)}{4m^2}\,
  \left[\frac{1}{2}\, S ( S + 1 ) - \frac{3}{4}\right]
\ ,
\label{mass}
\end{eqnarray}
where $\omega (n,\ell)$ and $\tau (n,\ell)$ are given in terms of
matrix elements involving the Lagrangian.

Spin-dependent relativistic corrections can be systematically studied
in the Wegner--Wilson loop approach. Their general structure
up to order $1/m^2$
(under the assumption of a fixed background field $A_\mu(x)$, cf.\
ref.\ \cite{Gr91})
has been obtained by Eichten and Feinberg (EF) \cite{EF81}
\begin{eqnarray}
V(r) & = & V_0(r) + \frac{V_0(r)' + 2V_1(r)' + 2V_2(r)'}{2 m^2 r}\;
  \vec{L} \cdot \vec{S} + \frac{V_4(r)}{3 m^2}\;
\vec{s}_1 \cdot \vec{s}_2
\nonumber\\  & & \quad
 + \frac{V_3(r)}{m^2} \left( \frac{\vec{r}\cdot \vec{s}_1 \;
  \vec{r}\cdot \vec{s}_2}{r^2} - \frac{1}{3}\, \vec{s}_1 \cdot \vec{s}_2
  \right)
\nonumber\\
 & & \quad + \;\; \hbox{spin-independent}\;\;\hbox{corrections}\ ,
\label{EFpotential}
\end{eqnarray}
where $V_0(r)$ is the spin-independent potential and $V_i(r)$ are related
to expectation values of the colour-electric and magnetic fields, e.g.\
($\vec{r} = \vec{r}_1 - \vec{r}_2$ and the limit $T\rightarrow
\infty$ is understood):
\begin{equation}
\left( \frac{r_i r_j}{r^2} - \frac{\delta_{ij}}{3} \right)\, V_3(r)
 + \frac{\delta_{ij}}{3}\, V_4(r) = \frac{g^2}{T}\,
  \int_{-T/2}^{T/2}\; \int_{-T/2}^{T/2}\; \left<
  B_i(\vec{r}_1,t) \, B_j(\vec{r}_2,t') \right> \mbox{d}t\,
  \mbox{d}t'\; / \left< 1 \right>
\ .
\label{Vfour}
\end{equation}
While (\ref{EFpotential}) is exact for the spin-dependent relativistic
corrections (for equal masses), the question of the spin-independent
relativistic corrections appears not to be yet completely understood
at the present time (for a review see e.g.\ \cite{Gr91}).

The potentials $V_i(r)$ are usually assumed to be flavour-independent.
This assumption is not only in agreement with observed mass splittings
of charmonium and bottomonium, but also confirmed by lattice-QCD
studies (for a recent study see e.g.\ ref.\ \cite{Bo94}).
The flavour dependence of mass splittings in the EF formulation
(\ref{EFpotential}) enters through the explicit $1/m^2$ factors,
but also through the mass dependence of the expectation values
\begin{equation}
 \langle r^q\, V_i^{(p)}(r) \rangle_{nl} =
  \int_0^\infty \mbox{d}r\; r^2\, R^2_{nl}(r)\; r^q\, V_i^{(p)}(r)
\ ,
\label{radialexp}
\end{equation}
where $R_{nl}(r)$ are the radial wave functions (normnalized to
$\int_0^\infty \mbox{d}r\ r^2\ R^2_{nl}(r) = 1$).
Our result (\ref{masssplit},\ref{mass}) has a structure similar to that
of (\ref{EFpotential}). For example, we may identify
\begin{equation}
  \tau(n,l) = \langle \frac{3}{4}\, V_4(r) \rangle_{nl}
\ .
\label{indentifi}
\end{equation}
On the other hand, it is straightforward to include
one-loop corrections to our result (\ref{masssplit}), while
the EF representations seem to be non-renormalizable objects, and hence
any attempt to calculate them in one-loop order should end up with
a divergent result \cite{Gr91}.

Although the functions $\omega (n,\ell)$ and $\tau (n,\ell)$ are matrix
elements of flavour-independent operators, they in fact depend on the
flavour through the states. We may obtain some idea on the flavour
dependence of $\omega (n,\ell)$ and $\tau (n,\ell)$ by comparing the
spectra of  bottomonium and charmonium.
For example,
the spin-orbit term (proportional to $\vec{L}\cdot\vec{S}$ in
(\ref{EFpotential})) and the tensor interaction
(last term in (\ref{EFpotential}))
do not contribute to either of the mass differences
$m(2{}^3S_1) - m(1{}^3S_1)$ or $\bar{m} (1{}^3P_J) - m(1{}^3S_1)$,
where $\bar{m} (1{}^3P_J)$ denotes the centre of gravity of the
three ${}^3P_J$ ($J=0,1,2$) states. Experimentally these mass
differences are $563\,$MeV ($589\,$MeV) and
$440\,$MeV ($428\,$MeV), respectively, for bottomonium (charmonium).
Hence, certainly
$\widetilde{\Lambda}(n,l)$ and $\tau(n,l)$
are basically flavour-independent.
%

Thus the parameters entering (\ref{mass}) seem
to be only weakly dependent on the heavy quark
flavour for the observable states of heavy quarkonia,
although heavy-flavour symmetry is not present any more once we include
subleading terms into ${\cal L}_0$.
However, it is
generally believed that in the heavy-mass limit
a quarkonium should behave as an almost  Coulombic system. This would
imply that $\widetilde\Lambda (n) \sim m$, $\omega (n) \sim m^2$ and
$\tau (n) \sim m^3$, which is not compatible with the data from
charmonium and  bottomonium.
(For (\ref{EFpotential}), the  vector-Coulomb case means spin-independent
terms are proportional to $\alpha_s^2 m$, while all spin-dependent
terms are $\propto \alpha_s^4 m$.)
This could mean that either these systems
are not close enough to the heavy-mass limit to become  Coulombic, or
that the heavy-mass limit is not the  Coulombic one.
Let us recall that the
heavy-light systems with a $b$ quark and to a lesser extent the ones with
a $c$ quark already behave as one would expect in the heavy-mass limit.
This clearly indicates that the flavour dependence of the matrix elements
parametrizing the long-distance effects
is smaller than suggested by the  Coulombic limit.

\section{Some Phenomenology}
In this section we shall consider a few applications of the formalism
outlined above. We shall not give a full discussion of the
phenomenological applications of the heavy-mass expansion
for heavy quarkonia,
but rather study a few simple examples in some detail,
based on the matching
calculation performed in section 5.

Compared to the conventional approach, and also compared to the
$v/c$ expansion advocated by BBL \cite{BB94},
we have a much
larger list of parameters, although spin symmetry reduces this number
to some extent. In a given order of the $1/m$ expansion they are all
of the same dimension and there is no a priori reason why some of
them should be less important than some others; at least the $1/m$
expansion does not give any hint.

The most simple example (although maybe academic) is the
semi-inclusive decays
of a heavy quarkonium into $e^+ e^-$ and light hadrons.
In leading order
of the expansion we have that $\eta_Q \to e^+ e^-$
light hadrons vanishes,
while we have for the $\psi_Q$
\begin{equation} \label{gex10}
\frac{1}{2}\,
\Gamma_0 (\psi_Q \to e^+ e^- \mbox{ light hadrons} ) =
{\cal C}^{ee}(A_2^{(1)},\mu ) \langle \psi | A_2^{(1)} | \psi \rangle
=  \frac{ 4 \pi}{3 m^2} \alpha^2 c_Q^2  a_0^{(1)}
\end{equation}
with
\begin{equation}
{\cal C}^{ee}(A_2^{(1)},\mu ) =-\frac{\pi}{3 m^2} \alpha^2 c_Q^2
\ .
\end{equation}

There are no corrections of order $1/m$, since there are no local
dimension-seven operators having non-vanishing matrix elements,
and also the
only non-local term, which is an insertion of a first-order
chromomagnetic
moment operator $G_1^{(\pm)}$, vanishes due to spin symmetry.

The first subleading contribution appears at order $1/m^2$. Using the
matching calculation of section 5 and taking into account the running
considered in section 6, we obtain
\begin{eqnarray}
\frac{1}{2}\, \Gamma_2  &=&
{\cal C}^{ee} (B_3^{(1)},\mu ) \langle \psi | B_3^{(1)} |
\psi \rangle_\mu
+{\cal C}^{ee} (D_3^{(8)},\mu ) \langle \psi | D_3^{(8)} |
\psi \rangle_\mu
\nonumber \\
&& + {\cal C}^{ee} (E_2^{(1)}, \mu )
                   \langle \psi | E_2^{(1)} | \psi \rangle_\mu
+  {\cal C}^{ee}(E_3^{(1)}, \mu ) \langle \psi | E_3^{(1)} |
\psi \rangle_\mu  \nonumber \\
&&
+{\cal C}^{ee} (E_4^{(1)},\mu ) \langle \psi | E_4^{(1)} |
\psi \rangle_\mu
+ {\cal C}^{ee}  (A_2^{(1)},\mu ) (k_2^{(1)}  + G^{(1)} + F^{(1)})
\end{eqnarray}
where $\mu$ is the renormalization point $\mu < m$,
and $F^{(1)}$ is the contribution of the single insertion of the
purely gluonic piece ${\cal L}_{glue}$ of the $1/m^2$ Lagrangian.

The  coefficients are given by
\begin{eqnarray}
{\cal C}^{ee}(B_3^{(1)},\mu ) &=& -\frac{\pi}{3 m^2} \alpha^2 c_Q^2
\nonumber \\
{\cal C}^{ee} (D_3^{(8)} ) &=& -\frac{ \pi}{3 m^2} \alpha^2 c_Q^2
\frac{16}{33 -2n_f} \ln \eta
\nonumber \\
{\cal C}^{ee} (E_2^{(1)}, \mu  ) &=&
{\cal C}^{ee} (E_3^{(1)}, \mu  ) =  -\frac{\pi}{12 m^2} \alpha^2 c_Q^2
\nonumber \\
{\cal C}^{ee} (E_4^{(1)}, \mu  ) &=&  -\frac{\pi}{3 m^2} \alpha^2 c_Q^2
\left[ \frac{1}{2} - \frac{8}{33 -2n_f} \ln \eta \right]
\ .
\end{eqnarray}
Note that we have an additional factor $1/(2m)^2$ in
front of the second-order contribution according to our definition
(\ref{gamexp}).

This expression looks relatively complicated, but it simplifies somewhat
due to spin symmetry. The matrix elements of the local operators are
given in terms of the three parameters $b_0^{(1)}$, $e_0^{(1)}$
and $d_0^{(8)}$, while the non-local terms introduce another
three parameters:
$k_2^{(1)}$, $G^{(1)}$, and $F^{(1)}$. At order $1/m^2$ there are thus
six parameters describing the decay rate, which may then be expressed
as
\begin{eqnarray}
\frac{1}{2}\, \Gamma_2  &=& \frac{\pi}{3 m^2} \alpha^2 c_Q^2 \left[
12 b_0^{(1)} + 8 e_0^{(1)} \left( 1 - \frac{12}{33 -2n_f}
\ln \eta \right)
\right.
\nonumber \\
&& \qquad \qquad \left. + 12 d_0^{(8)} \frac{16}{33 -2n_f} \ln \eta
 - k_2^{(1)} - G^{(1)} - F^{(1)} \right]
\label{above}
\ .
\end{eqnarray}

A further simplification can only be achieved with additional
theoretical prejudices.
A popular assumption (although ad hoc) for non-leptonic inclusive
decays of heavy-light mesons is vacuum insertion. Note, however,
that this is not a scale-invariant statement.
Usually  vacuum insertion is applied at some small scale, where
the matrix elements are estimated by e.g.\ a wave-function model.
There the local contributions are related
to the wave function and its derivatives at
the origin $\vec{x} = \vec{0}$, and the non-local contributions
correspond
to corrections to the wave function.
The application of vacuum insertion to the present case of
heavy-quarkonium decays
is quite a strong assumption, because it removes
already all the operators with a RCM derivative.
This follows from the fact that, with this assumption,
the c.m.s.\ frame of the hard annihilation process has to be the same
as the one of the heavy quarkonium. Furthermore, this assumption
leads to vanishing matrix elements for all colour $T^a \bigotimes T^a$
operators at the small scale $\mu$.
And finally, in such a picture
the contributions of the purely gluonic piece (\ref{Lglue}) vanish
(e.g.\ the term $F^{(1)}$ in (\ref{above})).

If we nonetheless use vacuum insertion at a small scale $\mu$ we find
\begin{equation}
\Gamma = \frac{ \alpha^2 c_Q^2}{m^2}\; \left[
\frac{8\pi}{3}\, a_0^{(1)}
 + \frac{4 \pi}{3 m^2}\, e_0^{(1)}\,
       \left( 1 - \frac{12}{33 -2n_f} \ln \eta \right)
  - \frac{\pi}{6 m^2}\, \left( k_2^{(1)} + G^{(1)} \right) \right]
\ .
\label{psiincl}
\end{equation}
The first term is the one familiar from non-relativistic
potential models:
Identifying
\begin{equation}
  a_0^{(1)} = \frac{3}{2}\, \left|\Psi(0)\right|^2 = \frac{3}{8\pi}\,
    \left|R(0)\right|^2
\end{equation}
we recover the well-known Royen--Weisskopf formula
($M=M({}^3S_1) \approx 2m)$:
\begin{equation}
  \Gamma({}^3S_1 \rightarrow \ell \bar{\ell}) =
          \frac{4\alpha^2c_Q^2}{M^2}\;  \left|R(0)\right|^2
\ ,
\label{Royenformula}
\end{equation}
which holds for the exclusive decay. However, it is this expression
we obtain to leading order also for the inclusive decay;
this indicates
that the exclusive mode $\psi \to e^+ e^-$
will saturate a large portion of the inclusive decay
$\psi \to e^+ e^- +$ light hadrons.

The non-logarithmic part of the $e_0^{(1)}$
$1/m^2$ correction in (\ref{psiincl}) has
also been discussed by BBL \cite{BB94} and by
Keung and Muzinich (KM) \cite{KM83}. It is in fact the
only correction that occurs in either approach.
BBL propose to identify
(the analogon to) $e_0^{(1)}$ with the limit as $r\rightarrow 0$
of $-\vec{\Delta}^2 R(r)$ with appropriate regularization. KM
calculate kinematical corrections to the leading-order amplitude.
They denote by $\epsilon$ the binding energy and
by $\vec{p}$ the relative
three-momentum of the heavy quark and antiquark. Then they evaluate
the amplitude at $\vec{p}\,{}^2 = m \epsilon$ rather than at
$\vec{p}\,{}^2=0$, as is the usual wave-function prescription
for $S$-wave decays. Their corrections proportional to $\epsilon/m$
are thus $\vec{p}\,{}^2/m^2$ corrections and can be identified
with those arising from the $E$-operators in our approach.

Let us emphasize that
the appearance of the logarithm in (\ref{psiincl}) indicates the
breakdown of the na\"{\i}ve potential-model calculations
also for $S$-wave
decays as was conjectured a long time ago \cite{BG76}.
Expanding
\begin{equation}
 \frac{6}{33 - 2n_f}\, \ln\eta \approx \frac{\alpha_s(m)}{\pi}\;
  \ln\frac{m}{\mu}
\ ,
\label{lnexpand}
\end{equation}
we obtain a contribution to the decay width proportional to
$|R''(0)|^2\, \ln(m/\mu)$. This manifestly violates the potential-model
ansatz that the infrared dynamics is given by the
meson's non-relativistic wave function
(and its derivatives), while the short-distance part is free of
infrared singularities.

As a further example, let us study the
hadronic decay of spin-triplet $P$-wave
quarkonia $n{}^3P_J = \chi_{QJ}(nP)$.
In the wave-function approach, the leading-order (in $\alpha_s(m)$)
decay is into two gluons for the $J=0,2$ states,
$\Gamma({}^3P_{0,2}\to g g) = O(\alpha_s^2)$, while the
$J=1$ state can first decay at $O(\alpha_s^3)$ into either three gluons
or a quark-antiquark-pair plus a gluon.
Moreover, it was found \cite{BG76,BG81,KK79}
that the quark-antiquark-gluon cuts are
singular in the limit of zero binding energy $\epsilon$
\begin{equation}
 \Gamma({}^3P_J \rightarrow q\bar{q}g) = \frac{n_f}{3}\,
  \frac{128}{3\pi}\, \frac{\alpha_s(m)^3}{M^4}\, \ln\frac{m}{\epsilon}
 \; \left| R'_{1P}(0) \right|^2
\ .
\label{singres}
\end{equation}
The magnitude of the logarithm is usually estimated by identifying
$1/\epsilon$ with the average radius of the ${}^3P$ states.
Again, the presence of an infrared sensitive logarithm signals
the breakdown of the usual factorization assumption that is behind
the wave-function approach. It has been argued that
the decay into $q \bar{q}g$ has to be considered as being of the
same perturbative order $O(\alpha_s^2)$ as the two-gluon decay since
the expresion $\alpha_s \ln(m/\epsilon) R'(0)^2$
has to be considered as a new
nonperturbative parameter besides $R'(0)^2$ describing
$P$-wave decays \cite{BB92,Sc94}.

The present formalism reproduces this result by generating the large
logarithm through the renormalization group running.
In fact, formally the decay into a pair of quarks is the
dominant one, since it is of the same order in $\alpha_s$
as the two-gluon decay but logarithmically enhanced,
see (\ref{pwavevac}) below. To see how this comes about
consider the $q\bar{q}$ decays of the spin-triplet
$P$-wave states in more detail. At the matching scale $m$ only
colour-octet contributions are present (section~5) and hence
the leading-order result is
\begin{equation} \label{gex80}
\frac{1}{2}\,
\Gamma_0 ( {}3 \chi  \to q \bar{q} \to \mbox{ light hadrons} ) =
-\frac{\pi}{6 m^2} \alpha_s^2 (m) n_f
\langle \psi | A_2^{(8)} | \psi \rangle
=  \frac{ 2 \pi}{3 m^2} \alpha_s^2 (m) n_f \,  a_0^{(8)}
\ .
\end{equation}
For the second-order contributions we find
\begin{eqnarray}
\frac{1}{2} \,
\Gamma_2  &=&
+{\cal C}^{qq} (B_3^{(8)},\mu ) \langle \psi | B_3^{(8)} |
 \psi \rangle_\mu
+{\cal C}^{qq} (D_3^{(8)},\mu ) \langle \psi | D_3^{(8)} |
\psi \rangle_\mu
+{\cal C}^{qq} (D_3^{(1)},\mu ) \langle \psi | D_3^{(1)} |
\psi \rangle_\mu
\nonumber \\
&& + {\cal C}^{qq}(E_2^{(8)}, \mu )
     \langle \psi | E_2^{(8)} | \psi \rangle_\mu
+  {\cal C}^{qq}(E_3^{(8)}, \mu )
\langle \psi | E_3^{(8)} | \psi \rangle_\mu
\nonumber \\
&&
+{\cal C}^{qq} (E_4^{(8)},\mu ) \langle \psi | E_4^{(8)} |
\psi \rangle_\mu
+ {\cal C}^{qq} (A_2^{(8)},\mu ) (k_2^{(8)} + G^{(8)}+ F^{(8)})
\ .
\end{eqnarray}
The  coefficients of the local operators are (at a scale $\mu < m$):
\begin{eqnarray}
{\cal C}^{qq}(B_3^{(8)},\mu ) &=& -\frac{\pi}{6 m^2}
\alpha^2_s (m) n_f
\left(1-  \frac{24}{33 -2n_f} \ln \eta   \right)
\nonumber \\
{\cal C}^{qq} (D_3^{(8)} ) &=& -\frac{10 \pi}{9 m^2} \alpha^2_s (m) n_f
                                \frac{1}{33 -2n_f} \ln \eta
\nonumber \\
{\cal C}^{qq} (D_3^{(1)} ) &=& -\frac{16 \pi}{27 m^2} \alpha^2_s (m) n_f
                                \frac{1}{33 -2n_f} \ln \eta
\nonumber \\
{\cal C}^{qq} (E_2^{(8)}, \mu  ) &=&
{\cal C}^{qq} (E_3^{(8)}, \mu  ) =  -\frac{\pi}{24 m^2} \alpha^2_s (m)
\nonumber \\
{\cal C}^{qq} (E_4^{(8)}, \mu  ) &=&  -\frac{\pi}{12 m^2}
\alpha^2_s (m)2
\left(1 - \frac{28}{3} \frac{1}{33 -2n_f} \ln \eta \right)
\ .
\end{eqnarray}
It is interesting to note that the renormalization group flow induces
operators ($D_3^{(8)}$ and $ D_3^{(1)}$) which have not been present at
the matching scale.
We may express the local contributions to the second-order
contribution in terms of the parameters
$b_1^{(8)}$, $d_1^{(8)}$, $d_1^{(1)}$, and $e_1^{(8)}$:
\begin{eqnarray}
\frac{1}{2}\,
\Gamma_2  &=&  \frac{\pi}{6 m^2} \alpha^2_s (m) n_f \left[
12 b_1^{(8)} \left(1-  \frac{24}{33 -2n_f} \ln \eta   \right)
+ 8 e_1^{(8)} \left(1 -  \frac{7}{33 -2n_f} \ln \eta   \right) \right.
\nonumber \\
&& \qquad + \left. \frac{128}{3} d_1^{(1)}
\frac{1}{33 -2n_f} \ln \eta
+ 80 d_1^{(8)} \frac{1}{33 -2n_f} \ln \eta
- k_2^{(8)} - G^{(8)} - F^{(8)} \right] .
\label{sublead}
\end{eqnarray}
We note that not only the leading contribution (\ref{gex80}), but
also the first subleading one (\ref{sublead})
is the same for all three states
${}^3 P_0$, ${}^3 P_1$, and ${}^3 P_2$. Thus, if the channel
involving a
$q \bar q$-pair really dominates
the decays of these states into light hadrons due
to the logarithmic enhancement, we have
\begin{equation}
\Gamma ({}^3 P_2 \to \mbox{ light hadrons})
=   \Gamma ({}^3 P_1 \to \mbox{ light hadrons})
=   \Gamma ({}^3 P_0 \to \mbox{ light hadrons})
\ .
\end{equation}

The second-order contribution (\ref{sublead})
is given in terms of seven parameters, but
this number reduces once vacuum insertion is assumed. Then only one
of them is non-zero, namely the parameter $d_1^{(1)}$, and the rate
takes the simple form
\begin{equation} \label{pwavevac}
\Gamma ({}^3 \chi_Q \to q \bar q (g) \to \mbox{ light hadrons } )
  =  \frac{32 \pi}{9 m^4} \alpha^2_s (m) n_f
     d_1^{(1)} \frac{1}{33 -2n_f} \ln \eta
\ .
\end{equation}
Furthermore, in a wave-function model the parameter $ d_1^{(1)}$
is given in terms of the derivative of the wave function at the
origin
\begin{equation}
 d_1^{(1)}(n) = \frac{3}{2\pi}\, \left| R'_{nP}(0) \right|^2
\ ,
\end{equation}
and if one expands (\ref{pwavevac}) again in
powers of $\alpha_s (m)$, see (\ref{lnexpand}),
we reproduce the result (\ref{singres}).

\section{Comparison with previous approaches and conclusion}
Traditionally \cite{No78} inclusive hadronic (and electromagnetic)
decays of heavy quarkonia are calculated
in the framework of the Bethe--Salpeter description of the bound state.
Assuming an instantaneous potential and working in the extreme
non-relativistic limit,
the decay width of a heavy quark-antiquark bound state
into light hadrons (l.h.) is written as
\begin{equation}
  \Gamma(n\,{}^{2S+1}L_J \rightarrow \mbox{l.h.}) = G(n)\;
  \hat{\Gamma}(Q\bar{Q}({}^{2S+1}L_J) \rightarrow \mbox{partons})
\ .
\label{naivedecay}
\end{equation}
This ansatz of separation into a short-distance part
$\hat{\Gamma}$ describing the decay of a free (unbound) $Q\bar{Q}$-pair
to decay into partons ($q\bar{q}$-pairs and gluons)
and a long-distance one $G(n)$ representing the non-perturbative
bound-state formation, is
motivated by the observation that the problem involves
two widely separated scales, the $Q\bar{Q}$ radius of the order
of $1\,$fm and the heavy-quark Compton wavelength $\lambda_{Q} \sim
1/m \ll r_{Q\bar{Q}}$. The non-perturbative part $G(n)$ is
expressed in terms of the non-relativistic
(Schr\"{o}dinger) wave function at zero relative coordinate $R_{nS}(0)$
(for $S$-wave decays, and the derivative $R'_{nP}(0)$
for $P$-wave decays), which
is calculated in a potential model or extracted from data.
The factor $G(n)$ thus depends
explicitly on the binding energy $\epsilon$. The $Q\bar{Q}$ decay,
on the other hand, is governed by a scale of the order of $\mu \sim m$,
and can hence be expanded in a series of $\alpha_s(m)$.
The decay rate is called factorizable into a long- and a
short-distance contribution \cite{BG76} if
the short-distance factor $\hat\Gamma$ is
calculable without encountering infrared divergencies.

Let us elaborate on the assumptions of the conventional
approach in some more detail.
Consider the transition amplitude ${\cal M}(J^P,P;k_i)$ of a
$J^P$ quark-antiquark bound state of mass $M$
(and four-momentum $P=(M,\vec{0})$)
into partons (light $q\bar{q}$-pairs and gluons) of four-momenta $k_i$
in terms of which the width is given by
$\Gamma \sim \left|{\cal M}\right|^2\,\mbox{dPS}$.
(For illustration we suppress colour and polarization indices.)
Introducing the Bethe--Salpeter bound-state wave function
$\Phi(J^P,P;q)$, dependent on the relative
quark and antiquark momenta $q$ ($p_{Q,\bar{Q}} = \frac{1}{2}P \pm q$),
the transition amplitude is written in this picture as
\begin{equation}
{\cal M} =  \int\frac{\mbox{d}^4 q}{(2\pi)^4}
  \mbox{Tr}\, \Phi(P,q)  {\cal O}(P,q,k_i)
\label{BSdecay}
\end{equation}
such that ${\cal O}(P,q,k_i)$ is the amplitude of a free (unbound)
$Q\bar{Q}$-pair to decay into light hadrons.
Upon a non-relativistic reduction
\begin{equation}
\Phi(P,q) = 2\pi\delta(q^0) \sum_{m,S_z} \psi_{nlm}(\vec{q})
  \left< lmS S_z\left| J J_z \right. \right> P_{SS_z}(P,q)
\ ,
\end{equation}
the amplitude becomes
\begin{equation}
 {\cal M} =
    \sum_{m,S_z} \int\frac{\mbox{d}^3q}{(2\pi)^3}
  \psi_{nlm}(\vec{q})
\;
\underbrace{
 \left< lmS S_z\left| J J_z \right.\right>
   \mbox{Tr} P_{SS_z}(P,q)
   {\cal O}(P,q,k_i) }_{ A(\vec{q}\,{}^2,\epsilon)}
\label{Mdef}
\end{equation}
where $p_i^2 = M^2/4 - \vec{q}\,{}^2
   = m^2 + \epsilon m - \vec{q}\,{}^2$,
$\epsilon$ is the binding energy, and $P_{SS_z}$ is a
projection operator.
The Schr\"odinger wave function in coordinate space extends over
distances of the order of the Bohr radius.
Correspondingly, the Fourier
transform  $\psi_{nlm}(\vec{q}\ )$ is non-zero only for
$|\vec{q}\, |/m \ll 1$, and hence $|\vec{q}\, |/m$
becomes a reasonable
expansion parameter.
For $S$-wave decays, the leading term of this expansion yields
the usual expression
for the decay amplitude proportional
to the non-relativistic wave function calculated at the origin in the
relative coordinate space:
\begin{equation}
{\cal M} \approx  \frac{1}{\sqrt{4\pi}} R_{nS}(0)
    A(\vec{q}\,{}^2=0,\epsilon)
\ .
\end{equation}
Here $R_{nl}(0) =  4\pi \psi_{nl}(0)$ ($l=S,P,D$, etc.)
is the radial wave function at zero distance $r=0$.
For $P$-wave decays, the terms linear in $\vec{q}$ in the expansion
of ${\cal O}$, as well as terms linear in $\vec{q}$ coming from
the small components of the relativistic wave function,
must be retained
yielding a final expression for the amplitude ${\cal M}$
proportional to the derivative of the wave function
for the $l=P$ state at zero $r$, $R'_{nP}(0)$.
Factorization is now said to hold if the limit $\epsilon
\rightarrow 0$ exists for the amplitude $A$.

Within this potential-model approach,
next-to-leading-order (NLO) perturbative
QCD corrections have been calculated.
Both the NLO corrections to ${}^1S_0$ (i.e.\ $\eta_{c}$) \cite{Ba79}
and ${}^3S_1$ (i.e.\ $J/\psi$) \cite{Ma81} decays
indeed obey the factorized form (\ref{naivedecay}).
However, there are three observations from which it becomes
obvious that this picture is too simple:
\begin{itemize}
\item Infrared sensitive logarithms $\sim \ln(m/\epsilon)$
appear in the calculation of $P$-wave decays, to be precise in
the NLO corrections to ${}^3P_{0,2}$ decays \cite{BG81} and already
in the leading-order expressions of ${}^1P_1$ and ${}^3P_1$
decays \cite{BG76}.
That is, without keeping the binding energy non-zero, the
perturbative part of the calculation would diverge. The factorization
(\ref{naivedecay}) thus breaks down.
\item The description of $S$-wave decays based on the strict
non-relativistic limit is not in agreement with data. For example,
a reasonable $\alpha_s$ determination from the ratio
$\Gamma({}^3S_1\rightarrow \mbox{l.h.})/
\Gamma({}^3S_1\rightarrow \ell\bar{\ell})$
is possible only once
a (rather large) adhoc relativistic correction factor is applied.
Other failures are the photon spectrum of ${}^3S_1$ decays and
photo- and hadroproduction of $J/\psi$ \cite{Sc94}. In fact,
logarithmic infrared divergences, predicted to arise also in
relativistic corrections to $S$-wave decays a long time ago \cite{BG76},
show up, cf.\ (\ref{psiincl},\ref{lnexpand}).
Within the wave-function approach one can distinguish two sources
of relativistic corrections, namely corrections to
(i) the amplitude ${\cal O}$ and the (ii) the wave function
$\psi$, cf.\ (\ref{Mdef}).
Relativistic corrections to the wave function are very difficult to
access since at present no full analysis of the spin-independent
relativistic corrections to the potential exists,
cf.\ (\ref{EFpotential}).
Kinematical corrections of type (i) have been discussed by Keung and
Muzinich some time ago \cite{KM83} for $S$-wave decays and more recently
applied also to $J/\psi$ photo- \cite{Ju93} and
hadroproduction \cite{Sc94}.
The prescription is to evaluate
the amplitude $A(\vec{q}\,{}^2,\epsilon)$, (\ref{Mdef}),
at $\vec{q}\,{}^2
= m \epsilon$ rather than at $\vec{q}\,{}^2=0$. In this way certain
binding-energy corrections $\propto \epsilon/m =
\vec{q}\,{}^2/m^2$ are kept.
\item Sizeable, non-perturbative corrections have
been predicted \cite{Vo78}
for annihilation decays of heavy quarkonia, at least
for the charmonium
system. Such higher-twist
corrections could arise from non-zero condensates
(in particular, the gluon condensate) and/or colour-octet intermediate
states, e.g.\ via a non-perturbative colour-E1 transition of the
$J/\psi$ into a (coloured) $\chi_{cJ}$ state
${}^3S_1 \rightarrow {}^3P_J^{(8)} + g$, followed by a ``hard"
($\mu \sim m$) decay ${}^3P_J^{(8)} \rightarrow gg$.
\end{itemize}

In this paper we aim at a systematic, QCD-based treatment of
inclusive annihilation decays of heavy quarkonia.
In our approach, the factorization of long- and short-distance
contributions is well defined. Furthermore, ``genuine"
relativistic corrections,
i.e.\ the ones proportional to the relative velocity of the
heavy quarks in the conventional language,
and what is usually called non-perturbative corrections
come out to have the same origin, namely the higher-order terms of the
$1/m$ expansion. In fact, since we may shift certain contributions
from the operators into the states in this expansion,
there is no unique distinction between the two kinds of corrections.

To achieve such a systematic description of quarkonia decays
we formulate an
effective field theory for heavy quarkonia, which follows rigorously
from the QCD Lagrangian. As a first step we write down an operator
product expansion for the inclusive annihilation decays of
heavy quarkonia
in order to separate long- and short-distance contributions.
Herein the distance
scale is set by the Compton wavelength of the heavy quark.
This provides us with a systematic $1/m$ expansion of the
short-distance contributions with mass-independent operators.
As the next step we expand the heavy-quark fields $Q_v^{(\pm)}(x)$ and
the Lagrangian. In the wave-function picture, the
expansions of the states and of the kernel correspond to corrections to
the amplitude $A$ in (\ref{Mdef}), while time-ordered insertion
of the higher-order Lagrangian correspond to corrections of the
wave function.

In the set-up of an effective theory approach to heavy quarkonia
we now observe a crucial difference
compared to heavy-light systems: The static limit does not exist
for a heavy quarkonium state. Divergent phases appear in the
quark-antiquark sector of HQET once the velocities of the two
heavy quarks differ by an amount of only $\Lambda_{QCD} /m$.
These divergent phases
can be and have to be absorbed into the heavy-quark states.
Using reparametrization invariance, we have shown that
it is necessary and sufficient to include the $1/m$ kinetic term into
the leading-order Lagrangian
in order to consider two heavy quarks
moving with the same velocity. The dynamics defined by this
static-plus-$1/m$-kinetic-energy
Lagrangian already contains the divergent phases
(and the binding of the two heavy quarks) as
an infrared effect, and this is the physical reason why the limit
$v \to v'$ exists.
Since the redefined states
of a heavy quarkonium (seemingly) do not have a static limit, it is,
in contrast to heavy-light systems,
not possible to describe the full mass dependence of the relevant matrix
elements in a $1/m$ expansion. Although the
short-distance part may still be written as a $1/m$ expansion with
mass-independent operators
(once divergent imaginary parts have been shifted from
the operators into the states by a suitable redefinition),
the matrix elements of these operators
with the redefined states become mass-dependent due to the
mass dependence
of the states.

As a result of the mass dependence of the states, heavy-flavour
symmetry does not hold anymore.
However, judging from the spectra
and also from the widths of charmonium compared to bottomonium,
the flavour dependence of the matrix elements does not
seem to be very strong; at least it appears to be much
weaker than one would expect for a  Coulombic system.

On the other hand, spin symmetry
is still preserved since the extra (kinetic) $1/m$ term in
the leading-order
Lagrangian is spin-symmetric. Hence the
heavy quarkonia states described by our leading-order
Lagrangian have to
fall into degenerate spin symmetry quartets. Spin symmetry,
furthermore,
allows a restriction of the number of independent
parameters describing
matrix elements involving heavy quarkonia states.

We have applied our approach to inclusive heavy-quarkonia
decays into light
quarks and leptons. The effective theory machinery allows
us to calculate
the logarithmic dependence on the heavy mass $m$ by studying the
renormalization of the operators mediating the decay. The matching of
QCD to the effective theory is performed at the large scale $m$, where
the coefficients of the operator are determined by comparing the QCD
result with the effective theory. This determines the
initial conditions
of the operator coefficients. The renormalization group
of the effective
theory then allows the coefficients
to be run down to some smaller scale $\mu$.
We now see how the factorization assumption of
(\ref{naivedecay}) is generalized in a proper QCD treatment:
A given inclusive annihilation decay is, in general, the sum of various
contributions, each of which is the product of
a non-perturbative contribution,
parametrized as the matrix element of an operator renormalized
at a scale $\mu$,
and a Wilson coefficient evaluated at the same scale.

In addition to matrix elements of local operators also
non-local contributions appear. From the conventional point of view
these correspond to corrections to the wave functions of the states.
Among these non-local terms one may identify spin-symmetry breaking
corrections, which in the wave-function language generate differences
between, for example, the ${}^1 S_0$ and the ${}^3 S_1$ wave functions.

The rate to
leading order is given in terms of forward matrix elements of
dimension-six operators,
which at the one-loop level are scale-invariant, and hence
no logarithmic enhancement through terms as $\ln(m/\mu)$ is present.
Due to
symmetries, the leading-order contribution is in general
given in terms
of only two parameters ($a_\ell^{(1)}$ and $a_\ell^{(8)}$)
for a given
orbital angular momentum $\ell$
(and a given $n$) of the heavy quarkonium state.

The first
subleading contribution is suppressed by two powers of the
heavy mass and
involves forward matrix elements of dimension-eight operators
as well
as non-local contributions corresponding to corrections to
the states.
The number of dimension-eight operators is quite large, although
it is somewhat reduced by spin symmetry. In total, the number of
independent
parameters is in general still large, namely 14 ($b_\ell^{(C)}$,
$c_\ell^{(C)}$, $d_\ell^{(C)}$, $e_\ell^{(C)}$,
$k_2^{(C)}$, $G_\ell^{(C)}$
and $F_\ell^{(C)}$) for each heavy quarkonium
angular momentum $\ell$.

Intuitively one may expect that some of the parameters are smaller
than others. If one assumes vacuum insertion, then all
matrix elements
may be interpreted as wave functions and its derivatives
taken at the
origin. In this way one is led to assume that all the
colour combinations
$T^a \bigotimes T^a$ are suppressed compared to the
$1 \bigotimes 1$ operators.
Furthermore, the wave functions for $P$-wave states vanish at the
origin, and this suggests that the matrix elements of the
operators $D_n^{(1)}$ dominate for the $P$-wave states.
Let us emphasize, however, that our
analysis is independent of these assumptions.
Although our approach has many common features with the expansion
advocated by BBL \cite{BB94} it is
more general since we give the expansion before additional,
less rigorous
assumptions have been applied:
\begin{enumerate}
\item  Within a set of operators of a given dimension
we do not neglect those that are suppressed by powers of the
relative $Q\bar{Q}$ velocity $v/c$ in the wave-function language.
For instance, we do not neglect an operator involving
a gluon field strength $[iD, iD]$ in comparison to $(iD)^2$,
although the two operators correspond to different
powers of $v/c$ in the language of \cite{BB94}.
Such a procedure is adequate
for Coulombic systems but may be a bad approximation for charmonium
and bottom\-onium, which certainly do not behave Coulomb-like.
Yet, up to order $(\tilde\Lambda / m )^2$ no difference appears between the
two approaches, since the matrix elements of antisymmetric products
of covariant derivatives vanish due to spin symmetry.
\item
We do not restrict the heavy quarkonium to the leading Fock state.
Corresponding arguments are again based on counting of powers
of $v/c$: Using perturbation theory,
each additional gluon associated with the (assumed)
dominant $Q\bar{Q}$ pair is ascribed an extra power
of $v/c$ \cite{BB94}
via the identification $v \sim \alpha_s(m\, v)$,
valid for a colour-Coulomb potential. Although this estimate may be
underlined by the multipole expansion, we do not know of any rigorous
derivation.
\item The non-local contributions from the
time-ordered products with the Lagrangian in $\Gamma_2$ (\ref{second})
allows us to rigorously define the spin-symmetry
breaking terms that appear at
order $1/m^2$. Thus in principle we can calculate the coefficients
that appear in relations as $R_{\psi} = R_{\eta_c}[1 + O(v^2/c^2)]$,
although in practise additional unknown parameters enter, namely
$G^{(c)}_l$ for each $l$.
\end{enumerate}
If we apply these additional assumptions, a large number of our
non-perturbative parameters may be dropped,
and our approach yields the same result up to order
$(\tilde\Lambda / m)^2$ as the one of ref.\ \cite{BB94}.

In conclusion, the approach presented here is based on QCD
and hence provides a model-independent basis for the description of
heavy-quarkonium physics.
It allows us to separate long and short distances,
where the short-distance contribution may be evaluated perturbatively.
The long-distance part is parametrized in terms of
matrix elements, which
involve up to order $1/m^2$ operators of dimension six and dimension
eight. These matrix elements are beyond the effective theory
approach and have to be taken either from data, estimated via
non-perturbative methods such as sum rules, or they may eventually
be calculated from lattice QCD.

\clearpage
\centerline{\large \bf APPENDIX}
\appendix
\section{Spinorology}
Similar to the case of only upper components \cite{Ma94}, we also
have only four independent matrices once we project on upper and
lower components. The projectors are
\begin{equation}
P_+ = \frac{1}{2} (1+ \slash{v}) \quad P_+ = \frac{1}{2} (1- \slash{v})
\end{equation}
and we have the mapping of the sixteen Dirac matrices
\begin{eqnarray}
1 \longrightarrow P_+ P_-  = 0
&\qquad &
\gamma_\mu \longrightarrow P_+ \gamma_\mu P_-
\nonumber\\
\gamma_\mu \gamma_5 \longrightarrow P_+ \gamma_\mu \gamma_5 P_- =
P_+  \gamma_5 P_- v_\mu
&\qquad &
\gamma_5 \longrightarrow P_+ \gamma_5 P_-
\nonumber \\
(-i) \sigma_{\mu \nu} \longrightarrow P_+ (-i) \sigma_{\mu \nu}P_- =
P_+ (v_\mu \gamma_\nu - v_\nu \gamma_\mu)  P_-
\ .
\label{sigrep}
\end{eqnarray}
We chose the four matrices to be $P_+ \gamma_5 P_-$ (corresponding
to the
unit matrix) and $P_+ \gamma_\mu P_-$ (corresponding to the
three Pauli
matrices). Note that $v^\mu P_+ \gamma_\mu P_- = 0$, so these
are indeed
only three matrices. Hence, stricktly speaking, all Lorentz indices
in (\ref{As}) and (\ref{Bidef}--\ref{Eidef}) should be
written as perpendicular indices only defined by
\begin{equation}
 a_{\mu}^{\perp} = a_{\rho}\, \left( g^{\rho\mu} - v^{\rho}\, v^{\mu}
      \right) \ .
\label{defperp}
\end{equation}
Note that
\begin{equation}
  g^{\mu\nu}\; g_{\mu\nu}^{\perp} = 3 \ .
\label{gmumu}
\end{equation}

One may obtain the projections for any Dirac matrix $\Gamma$
in terms of
these four matrices by the trace formula
\begin{equation}
P_+ \Gamma P_- = \frac{1}{2}
                 \mbox{ Tr } \left\{\Gamma P_+ \gamma_5 P_-\right\}
                 P_+ \gamma_5 P_-
   + \frac{1}{2} \mbox{ Tr } \left\{\Gamma P_+ \gamma^\mu P_-\right\}
                 P_+ \gamma_\mu^{\perp} P_-
\ .
\label{project}
\end{equation}
The projections with $P_-$ and $P_+$ interchanged are
obtained by replacing
$v \to -v$ in the above equations.

\section{Matching calculation to $O(1/m^2)$}
In this appendix we collect a few useful relations that appear
frequently in the matching calculation.

\subsection{Expansions of bilinears}
Using (\ref{fexpFW}), (\ref{RCM}), and (\ref{RRM})
we obtain, for the bilinears
up to and including order $1/m^2$:
\begin{eqnarray}
\Qqp \Gamma \Qm & = & \hqp \Gamma \hm
\nonumber\\
 & + & \frac{1}{4m} \left(
 i\ \paperp_{\mu} \left( \hqp \left[ \Gamma,\gamma^{\mu}
\right] \hm \right)
+ \hqp \left\{ \Gamma,\Dperplr \right\} \hm \right)
\nonumber\\
 & + & \frac{1}{16 m^2} \left(
   i\ \paperp_{\mu} \, i\ \paperp_{\nu} \left( \hqp \gamma^{\mu}
  \Gamma \gamma^{\nu} \hm \right)
\right.
\nonumber\\ & &
   + i\ \paperp_{\mu} \left( \hqp \gamma^{\mu}
\Gamma \Dperplr \hm \right)
   - i\ \paperp_{\mu} \left( \hqp \Dperplr
\Gamma \gamma^{\mu} \hm \right)
\nonumber\\ & & \left.
 - \hqp \Dperplr\, \Gamma \Dperplr \hm \right)
   + O\left(\frac{1}{m^3} \right)
\ .
\label{biexpansion}
\end{eqnarray}
With the help of (\ref{project}) and the equations of motion
$i\ v\cdot \paperp \left( \hqp \Gamma \hm \right)
= 0 = \hqp\, v \cdot \Dperplr\  \Gamma \hm$
we find the following
expansions of the bilinears up to and including order $1/m^2$:
\begin{eqnarray}
\Qqp\, \Qm & = & \frac{1}{2\ m}\; A_6
\nonumber\\
\Qqp\, \gamma_5\, \Qm & = & A_1 + \frac{1}{16\ m^2}\left(
  - A_9 + 2\ A_{12} + A_{14} - A_{17} \right)
\nonumber\\
\Qqp\, \gamma^{\alpha}\, \Qm & = & A_2^{\alpha} + \frac{1}{2\ m} \;
  v^{\alpha}\, A_3
\nonumber\\ & &
       + \frac{1}{16\ m^2}\left(
  - A_{10}^{\alpha} + 2\ A_{11}^{\alpha}   - 2\ A_{13}^{\alpha}
  + A_{15}^{\alpha} - A_{16}^{\alpha} + A_{18}^{\alpha} \right)
\nonumber\\
\Qqp\, \gamma_5\, \gamma^{\alpha}\, \Qm & = & - v^{\alpha}\, A_1
  + \frac{1}{2\ m} \left(  A_4^{\alpha} + A_7^{\alpha} \right)
\nonumber\\ & &
 + \frac{1}{16\ m^2}\, v^{\alpha}\;
  \left( - A_9 + 2\  A_{12}   +  A_{14} - A_{17}\right)
\nonumber\\
\Qqp\, (-i)\ \sigma^{\alpha\beta}\, \Qm & = & v^{\alpha}\,
A_2^{\beta}
 - v^{\beta}\, A_2^{\alpha}
\nonumber\\ & &
  + \frac{1}{2\ m} \left(  A_5^{\beta\alpha} -A_5^{\alpha\beta}
  - A_8^{\alpha\beta} \right)
\nonumber\\ & & \hspace*{-2.5cm}
 + \frac{1}{16\ m^2}\left( v^{\alpha}\; \left( A_{10}^{\beta}
  - 2\  A_{11}^{\beta}  + 2\  A_{13}^{\beta} - A_{15}^{\beta}
     +  A_{16}^{\beta}  -  A_{18}^{\beta} \right)
  - (\alpha\leftrightarrow \beta) \right)
\ .
\label{QQbarexpansion}
\end{eqnarray}
Here we have introduced
\begin{eqnarray}
A_1 & = & \hqp\, \gamma_5\, \hm
\nonumber\\
A_2^{\alpha} & = & \hqp\, \gaperpal\, \hm
\nonumber\\
A_3 & = & i\, \paperpal\left(\hqp\, \gaperp_{\alpha}\, \hm\right)
\nonumber\\
A_4^{\alpha} & = & i\, \paperpal\left(\hqp\, \gamma_5\, \hm\right)
\nonumber\\
A_5^{\alpha\beta} & = & i\, \paperpal\left(\hqp\, \gaperpbe\,
\hm\right)
\nonumber\\
A_6 & = & \hqp\, \Dperplr\, \hm
\nonumber\\
A_7^{\alpha} & = & i\, \epsilon^{\alpha\mu\nu\rho}\, v_{\mu}\;
 \hqp\, \gaperp_{\nu}\, \Dperplrrho\, \hm
\nonumber\\
A_8^{\alpha\beta} & = & i\, \epsilon^{\alpha\beta\mu\rho}\,
 v_{\mu}\;
 \hqp\, \gamma_5\, \Dperplrrho\, \hm
\nonumber\\
A_9 & = & i\, \paperprh\; i\, \paperp_{\rho}\;
 \hqp\, \gamma_5\, \hm
\nonumber\\
A_{10}^{\alpha} & = & i\, \paperprh\; i\, \paperp_{\rho}\;
 \hqp\, \gaperpal\, \hm
\nonumber\\
A_{11}^{\alpha} & = & i\, \paperpal\; i\, \paperprh\;
 \hqp\, \gaperp_{\rho}\, \hm
\nonumber\\
A_{12} & = & i\, \epsilon^{\alpha\mu\nu\rho}\, v_{\alpha}\; i\,
\paperp_{\mu}\; \hqp\, \gaperp_{\nu}\, \Dperplrrho\, \hm
\nonumber\\
A_{13}^{\alpha} & = & i\, \epsilon^{\alpha\mu\nu\rho}\,
v_{\mu}\; i\,
 \paperp_{\nu}\; \hqp\, \gamma_5\, \Dperplrrho\, \hm
\nonumber\\
A_{14} & = & \hqp\, \Drhoperplr\, \gamma_5\, \Dperplrrho\, \hm
\nonumber\\
A_{15}^{\alpha} & = & \hqp\, \Drhoperplr\, \gaperpal\,
\Dperplrrho\, \hm
\nonumber\\
A_{16}^{\alpha} & = & \hqp\, \Dalperplr\, \Dperplr\, \hm
                   + \hqp\, \Dperplr\, \Dalperplr\, \hm
\nonumber\\
A_{17} & = & i\ \epsilon^{\alpha\rho\mu\sigma}\,
v_{\alpha}\; \hqp\,
  \Dperplrrho\, \gamma^{\perp}_{\mu}\, \Dperplrsig\, \hm
\nonumber\\
A_{18}^{\alpha} & = & i\ \epsilon^{\alpha\mu\rho\sigma}\,
v_{\mu}\; \hqp\,
  \Dperplrrho\, \gamma_5\, \Dperplrsig\, \hm
\ .
\label{Aidef}
\end{eqnarray}
\subsection{Expansions of quatrilinears}
With the help of the bilinears (\ref{Aidef}), we obtain
$1/m$ expansions of quatrilinears $A_i \otimes \{g,\epsilon\}
\otimes A_j^*$.
Owing to the fact that quatrilinears have to be parity-even,
and that
total derivatives may be neglected, we find that there are two
dimension-six operators, no dimension-seven operator, and $17$
dimension-eight operators.
Three more dimension-eight operators ($C_3+C_4$, $D_3 + D_4$,
$E_2 - E_3$) appear only if also the kernel is expanded.
Including colour, there are twice as many operators
\begin{equation}
 O_i^{(8)} = T^a \otimes T^a\, O_i^{(1)}
\ .
\label{Oieight}
\end{equation}
The dimension-six operators are
\begin{eqnarray}
- A_1^{(1)} & = & A_1   A_1^*
\nonumber\\
A_2^{(1)} & = & A_2^{\alpha}   A_{2\alpha}^*
\ .
\label{quatris}
\end{eqnarray}
The dimension-eight operators that can be constructed
from (\ref{Aidef}) are:
\begin{eqnarray}
B_1^{(1)} & = & A_4^{\alpha}   A^*_{4\alpha}
  = A_9   A^*_1
\nonumber\\
- B_2^{(1)} & = &  A_3   A^*_3
  = A_{11}^{\alpha}   A^*_{2\alpha}
\nonumber\\
- B_3^{(1)} & = &  A_5^{\alpha\beta}   A^*_{5\alpha\beta}
  = A_{10}^{\alpha}   A^*_{2\alpha}
\label{CMCM}
\\ \nonumber\\
C_1^{(1)} & = & \frac{1}{2}\ i\ \epsilon^{\alpha\beta\mu\nu}\,
v_{\mu}\,
     A_{8\alpha\beta}\, A_{4\nu}^*
\nonumber\\
 - C_2^{(1)} & = & A_6   A^*_3 + A_3   A^*_6
\nonumber\\
 C_3^{(1)} - C_4^{(1)} & = &  i\ \epsilon^{\alpha\beta\mu\nu}\,
v_{\mu}\,
     A_{5\alpha\beta}\, A_{7\nu}^*
\nonumber\\
 C_5^{(1)} & = & A_{12}   A^*_1 + A_1   A^*_{12}
  = - \left( A_7^{\alpha}   A^*_{4\alpha}
          + A_4^{\alpha}   A^*_{7\alpha}  \right)
\nonumber\\
 - C_6^{(1)} & = &  A_{13}^{\alpha}   A^*_{2\alpha}
          + A_2^{\alpha}   A^*_{13\alpha}
       = A_{8}^{\alpha\beta}   A^*_{5\alpha\beta}
\label{RECM}
\\ \nonumber \\
 - D_1^{(1)} & = & \frac{1}{2} A_8^{\alpha\beta}
 A^*_{8\alpha\beta}
\nonumber\\
 - D_2^{(1)} & = &  A_6   A^*_{6}
\nonumber\\
 D_3^{(1)} -  D_4^{(1)} & = &  A_7^{\alpha}   A^*_{7\alpha}
\nonumber\\
 D_5^{(1)} & = & - \frac{1}{2}\ i\ \epsilon^{\alpha\beta\mu\nu}\,
   v_{\mu}\, A_{8\alpha\beta}\, A^*_{7\nu}
 + \mbox{h.c.}
\label{RERED}
\\ \nonumber \\
 - E_1^{(1)} & = & A_{14}  A^*_{1} + A_{1}  A^*_{14}
\nonumber\\
E_2^{(1)} + E_3^{(1)} & = & A_{16}^{\alpha}   A^*_{2\alpha}
      + A_{2}^{\alpha}   A^*_{16\alpha}
\nonumber\\
E_4^{(1)} & = & A_{15}^{\alpha}   A^*_{2\alpha}
      + A_{2}^{\alpha}   A^*_{15\alpha}
\nonumber\\
E_5^{(1)} & = & A_{17}   A^*_{1}
      + A_{1}   A^*_{17}
\nonumber\\
 - E_6^{(1)} & = & A_{18}^{\alpha}   A^*_{2\alpha}
      + A_{2}^{\alpha}   A^*_{18\alpha}
\ .
\label{REREE}
\end{eqnarray}
\subsection{Two-component representation}
In the rest frame of the quarkonium we
have $v^{\mu} = (1,\vec{0})$.
Using
the Dirac representation of the gamma-matrices let us introduce
the two-component spinors $\psi$ and $\chi$ via
\begin{equation}
  \hp = \left( \begin{array}{c} \psi \\ 0 \end{array} \right)
\quad , \quad
  \hm = \left( \begin{array}{c} 0 \\ \chi \end{array} \right)
\ .
\label{psidef}
\end{equation}
Then we can express the bilinears as follows
\begin{eqnarray}
 A_1 & = & \psidag\, \chi
\nonumber\\
 \vec{A}_2 & = & \psidag\, \bfsigma\, \chi \quad (A_2^0 = 0)
\nonumber\\
 A_3 & = & - \bfpa\, \cdot \left( \psidag\, \bfsigma\, \chi \right)
\nonumber\\
 \vec{A}_4 & = & \bfpa\, \left( \psidag\, \chi \right)
 \quad (A_4^0 = 0)
\nonumber\\
 A_5^{ij} & = & i\ \partial^i\, \left( \psidag\, \sigma^j\,
\chi \right)
\nonumber\\
 A_6 & = & - \psidag\, \bfdif \cdot \bfsigma\, \chi
\nonumber\\
 A_7^i & = & - i \ \psidag \left( \bfsigma \times \bfdif
\right)_i\, \chi
\nonumber\\
 A_8^{ij} & = & i\ \epsilon^{ijk}\, \psidag\,
   i\ \stackrel{\longleftrightarrow}{D_k} \, \chi
\nonumber\\
 A_9 & = & - \left(\bfpa \right)^2\, \psidag\, \chi
\nonumber\\
 \vec{A}_{10} & = & - \left(\bfpa \right)^2\, \psidag\,
  \bfsigma\, \chi
\nonumber\\
 \vec{A}_{11} & = & - \bfpa \left(\bfpa \cdot \, \left(
  \psidag \bfsigma\, \chi \right) \right)
\nonumber\\
 A_{12} & = & i\ \epsilon^{ijk}\, i\ \partial_i\, \psidag\,
 \sigma_j\, i\ \stackrel{\longleftrightarrow}{D_k}\, \chi
\nonumber\\
 A_{13}^i & = & - i\ \epsilon^{ijk}\, i\ \partial_j\, \psidag\,
   i\ \stackrel{\longleftrightarrow}{D_k}\, \chi
\nonumber\\
 A_{14} & = & - \psidag \left( \bfdif \right)^2 \chi
\nonumber\\
 \vec{A}_{15} & = & - \psidag\, \bfsigma\,
  \left( \bfdif \right)^2 \chi
\nonumber\\
 \vec{A}_{16} & = & - \psidag \,
  \left[ \bfdif\, \left( \bfdif \cdot \bfsigma \right)
    + \left( \bfdif \cdot \bfsigma \right)\, \bfdif \right]\, \chi
\nonumber\\
 A_{17} & = & i\ \epsilon^{ijk}\, \psidag
  i\ \stackrel{\longleftrightarrow}{D_i}\,
  \sigma_j\, i\ \stackrel{\longleftrightarrow}{D_k}\, \chi
\nonumber\\
 A_{18}^k & = & - i\ \epsilon^{ijk}\, \psidag\,
 i\ \stackrel{\longleftrightarrow}{D_i}\,
 i\ \stackrel{\longleftrightarrow}{D_k}\, \chi
\ .
\label{Aitwo}
\end{eqnarray}
The complex-conjugate expressions are obtained from
(\ref{Aitwo}) by the
replacements $\psidag \rightarrow \chidag$, $\chi \rightarrow \psi$,
$i \rightarrow -i$ (using $\sigma_i^{\dagger} = \sigma_i$).
The quatrilinears are obtained using
$\epsilon_{ijk}\, \epsilon_{ijl} = 2\ \delta_{kl}$.
The connection to the operators of ref.\ \cite{BB94}
is then straightforward
($c=1$ or $8$)
\begin{eqnarray}
 A_1^{(c)} & = & - {\cal O}_c({}^1S_0)
\nonumber\\
 A_2^{(c)} & = & - {\cal O}_c({}^3S_1)
\nonumber\\
 \frac{1}{4}\, D_1^{(c)} & = & {\cal O}_c({}^1P_1)
\nonumber\\
 \frac{1}{4}\, D_2^{(c)} & = & 3\, {\cal O}_c({}^3P_0)
\nonumber\\
 \frac{1}{4}\, D_3^{(c)} & = & {\cal O}_c({}^3P_0)
      + {\cal O}_c({}^3P_1) + {\cal O}_c({}^3P_2)
\nonumber\\
 \frac{1}{4}\, D_4^{(c)} & = & {\cal O}_c({}^3P_0)
      - {\cal O}_c({}^3P_1) + {\cal O}_c({}^3P_2)
\nonumber\\
 \frac{1}{4}\, E_1^{(c)} & = &  2\, {\cal P}_c({}^1S_0)
\nonumber\\
 \frac{1}{4}\, E_4^{(c)} & = &  2\, {\cal P}_c({}^3S_1)
\nonumber\\
 \frac{1}{4}\, \left[ E_2^{(c)} +  E_3^{(c)} \right] & = &
   4\, {\cal P}_c({}^3S_1,{}^3D_1) + \frac{4}{3}\,
{\cal P}_c({}^3S_1)
\ .
\label{BBLconnection}
\end{eqnarray}

\subsection{Matching calculation}
The short-distance coefficients appearing in (\ref{Leading}) and
(\ref{second}) at the scale $\mu=m$ are obtained from the
annihilation part of the scattering amplitude ${\cal M}$ for
$Q\bar{Q} \rightarrow Q\bar{Q}$ computed in full QCD. The amplitude,
calculated in QCD perturbation theory for on-shell
quarks and antiquarks, is Taylor-expanded in the ``small" components
of the heavy quark momenta. The small component of a momentum
is defined as the component perpendicular to the direction of
the quarkonium momentum $Mv$. Denoting
the momenta of the initial $Q\bar{Q}$ pair
by $p$ and $\bar{p}$ for $Q$ and $\bar{Q}$, respectively,
and the ones of the final $Q\bar{Q}$-pair
by $p'$ and $\bar{p}'$ for $Q$ and $\bar{Q}$, respectively
we have:
\begin{eqnarray}
  p & = & v \sqrt{m^2 - p_{\perp}^2} + p_{\perp}
\nonumber\\
  & = & m v\left[1 - \frac{1}{2m^2}\,
       \left( P + \pi \right) ^2 + \ldots \right] + P + \pi
\nonumber\\
  \bar{p} & = & m v\left[1 - \frac{1}{2m^2}\,
       \left( P - \pi \right) ^2 + \ldots \right] +  P - \pi
\nonumber\\
    p' & = & m v\left[1 - \frac{1}{2m^2}\,
       \left( P + \eta \right) ^2 + \ldots \right] + P + \eta
\nonumber\\
  \bar{p}' & = & m v\left[1 - \frac{1}{2m^2}\,
       \left( P - \eta \right) ^2 + \ldots \right] + P - \eta
\ ,
\label{momexpand}
\end{eqnarray}
where $\pi^2 = \eta^2$.
Note that both the kernel and the spinors have to expanded
to the desired
order. The latter expansion follows from (\ref{fexpFW}).
For example, up to order $1/m^2$ we have
\begin{equation}
  Q^{(+)} = \left( 1 + \frac{P^2 + \pi^2}{8 m^2} \right) \,
    \left(1 + \frac{\slash{P} + \slash{\pi}}{2 m} \right) \; P_+
\ .
\end{equation}
The result of this procedure is then matched to the operators
in (\ref{Leading}) and (\ref{second}) by identifying the momenta
with the derivatives appearing in the operators
\begin{eqnarray}
  \bar{h}^{(+)} \, \Gamma\,
              i \stackrel{\longleftrightarrow}{D}_\mu \,
               h^{(-)}
  & \rightarrow & \bar{h}^{(+)} \, \Gamma\, \eta_{\mu}\, h^{(-)}
\nonumber\\
  \bar{h}^{(-)} \, \Gamma\,
              i \stackrel{\longleftrightarrow}{D}_\mu \,
               h^{(+)}
  & \rightarrow & \bar{h}^{(-)} \, \Gamma\, \pi_{\mu}\,  h^{(+)}
\nonumber\\
   i \partial_\mu
              \left(\bar{h}^{(\pm)}\, \Gamma\,  h^{(\mp)} \right)
  & \rightarrow &  \bar{h}^{(\pm)}\, \Gamma\,  P_\mu\, h^{(\mp)}
\ .
\label{substit}
\end{eqnarray}
Then the coefficients ${\cal C}(O_i^{(c)})$ of
(\ref{second}) can be read off
from the dimension-eight contribution to the amplitude
\begin{eqnarray}
{\cal M}_{d=8} & = & \left(\frac{1}{2m}\right)^2\,
\sum_{c=1,8}\, \left[
      \sum_{i=1}^3\, {\cal C}(B_i^{(c)},m)\, B_i^{(c)}
   +  \sum_{i=1}^6\, {\cal C}(C_i^{(c)},m)\, C_i^{(c)}
\right.
\nonumber\\
& & \left.
   +  \sum_{i=1}^5\, {\cal C}(D_i^{(c)},m)\, D_i^{(c)}
   +  \sum_{i=1}^6\, {\cal C}(E_i^{(c)},m)\, E_i^{(c)}  \right]
\nonumber\\
 & \equiv &  \frac{1}{4m^2}\, \sum_{c=1,8}\, X_c\, \left\{
  {\cal C}(B_1^{(c)},m)\, P^2\,  \gamma_5 \otimes \gamma_5
\right. \nonumber\\ & &
 + {\cal C}(B_2^{(c)},m)\, \Pslash \otimes \Pslash
 + {\cal C}(B_3^{(c)},m)\, P^2\, \gamma^{\alpha} \otimes
\gamma_{\alpha}
\nonumber\\
 & & + \, {\cal C}(C_1^{(c)},m)\, (\pi + \eta) \cdot P\,
  \gamma_5 \otimes \gamma_5
 + {\cal C}(C_2^{(c)},m)\,  \left[ \pislash \otimes \Pslash
       + \Pslash \otimes \etaslash  \right]
\nonumber\\
  & & + \, {\cal C}(C_3^{(c)},m)\,  (\pi + \eta) \cdot P\,
  \gamma^{\alpha} \otimes \gamma_{\alpha}
 + {\cal C}(C_4^{(c)},m)\, \left[ \Pslash \otimes \pislash
         + \etaslash \otimes \Pslash \right]
\nonumber\\ & & + \,
  {\cal C}(C_5^{(c)},m)\, (-i)\ \epsilon^{\alpha\beta\gamma\delta}\,
  v_{\alpha}\, P_{\beta}\, \left[
  \pi_{\gamma}\, \gamma_{\delta} \otimes \gamma_5 +
  \eta_{\gamma}\, \gamma_5 \otimes \gamma_{\delta} \right]
\nonumber\\ & & + \,
  {\cal C}(C_6^{(c)},m)\, (-i)\ \epsilon^{\alpha\beta\gamma\delta}\,
  v_{\alpha}\, P_{\beta}\, \left[
  \pi_{\gamma}\, \gamma_5 \otimes \gamma_{\delta} +
  \eta_{\gamma}\, \gamma_{\delta} \otimes \gamma_5 \right]
\nonumber\\
 & & + \, {\cal C}(D_1^{(c)},m)\, \pi\cdot\eta\, \gamma_5
 \otimes \gamma_5
 + {\cal C}(D_2^{(c)},m)\,  \pislash \otimes \etaslash
\nonumber\\ & & + \,
   {\cal C}(D_3^{(c)},m)\, \pi \cdot \eta\,
   \gamma^{\alpha} \otimes \gamma_{\alpha}
 + {\cal C}(D_4^{(c)},m)\, \etaslash \otimes \pislash
\nonumber\\ & & + \,
  {\cal C}(D_5^{(c)},m)\, (-i)\ \epsilon^{\alpha\beta\gamma\delta}\,
  v_{\alpha}\, \pi_{\beta}\, \eta_{\gamma}\,
 \left[ \gamma_5 \otimes \gamma_{\delta} + \gamma_{\delta} \otimes
  \gamma_5 \right]
\nonumber\\
 & & + \, {\cal C}(E_1^{(c)},m)\, \left[ \pi^2 + \eta^2 \right]\,
          \gamma_5 \otimes \gamma_5
 + \left\{ {\cal C}(E_2^{(c)},m) + {\cal C}(E_3^{(c)},m) \right\}\,
       \left[ \pislash \otimes \pislash
        + \etaslash \otimes \etaslash \right]
\nonumber\\ & & \left.
 + \, {\cal C}(E_4^{(c)},m)\, \left[ \pi^2 + \eta^2 \right]\,
          \gamma^{\alpha} \otimes \gamma_{\alpha}
        \right\}
\ ,
\label{dimeneight}
\end{eqnarray}
where we suppressed the spinors and abbreviated the colour
structure
\begin{equation}
 X_1 = 1 \otimes 1 \quad , \quad X_8 = T^a \otimes T^a
\ .
\label{Xdef}
\end{equation}
Note that $E_5$ and $E_6$ give only field-strength terms, i.e.\ vanish
for commuting Dirac matrices.
The dimension-six part is simply
\begin{eqnarray}
{\cal M}_{d=6} & = & \frac{1}{m^2}\, \sum_{c=1,8}\, \sum_{i=1}^{2}\,
  {\cal C}(A_i^{(c)},m)\, A_i^{(c)}
\nonumber\\
 & \equiv &  \frac{1}{m^2}\, \sum_{c=1,8}\, X_c\, \left\{
  {\cal C}(A_1^{(c)},m)\, \gamma_5 \otimes \gamma_5
 + {\cal C}(A_2^{(c)},m)\, \gamma^{\alpha} \otimes
\gamma_{\alpha} \right\}
\ .
\label{dimsenix}
\end{eqnarray}

\section{Operators in terms of quarkonia quantum numbers}
In order to calculate the hadronic matrix elements
using vacuum insertion,
it is convenient
to re-express the operators $B$, $C$ and $D$ in terms of other
operators in which for
both of the heavy-quark bilinears the
``orbital angular momentum'' corresponding to the
derivative is coupled
to the total spin of the quarks to some total angular momentum $J$.
If $\nabla _\mu$ is either the RRM or the RCM derivative
($i \stackrel{\longleftrightarrow}{D}_\mu$ or $i \partial_\mu$,
respectively), these couplings are
\begin{eqnarray}
{}^1 P_1 \otimes {}^1 P_1 & \widehat{=} &
          \left(\bar{h}^{(-)} \gamma_5 \nabla_\mu h^{(+)}\right)
          \left(\bar{h}^{(-)} \gamma_5 \nabla^\mu h^{(+)}\right)
\nonumber\\
{}^3 P_0 \otimes {}^3 P_0 & \widehat{=} &
      \frac{1}{3}\;
          \left( \bar{h}^{(-)} \gamma^\mu \nabla_\mu h^{(+)} \right)
          \left( \bar{h}^{(-)} \gamma^\nu \nabla_\nu h^{(+)} \right)
\nonumber\\
{}^3 P_1 \otimes {}^3 P_1 & \widehat{=} &
      \frac{1}{2}\;
          i \varepsilon_{\alpha \beta \mu \nu} v^\beta
          \left(\bar{h}^{(-)} \gamma^\mu \nabla^\nu h^{(+)} \right)
           i \varepsilon^{\alpha \beta' \mu' \nu'} v_{\beta'}
          \left(\bar{h}^{(-)} \gamma_{\mu'} \nabla_{\nu'}
 h^{(+)} \right)
\nonumber\\
{}^3 P_2 \otimes {}^3 P_2 & \widehat{=} &
          \left(\bar{h}^{(-)} \gamma^{[\mu} \nabla^{\nu]}
h^{(+)} \right)
          \left(\bar{h}^{(-)} \gamma_{[\mu} \nabla_{\nu]}
h^{(+)} \right)
\end{eqnarray}
where the bracket denotes the symmetric traceless combination of a
tensor (cf.\ (\ref{defperp},\ref{gmumu}))
\begin{equation}
  X_{\perp}^{[\mu\nu]} = \frac{1}{2}\, \left( X_{\perp}^{\mu\nu}
       + X_{\perp}^{\nu\mu} \right) - \frac{1}{3}\,
g_{\perp}^{\mu\nu}\,
    X_{\perp}^{\mu}{}_{\mu}
\ .
\end{equation}
The first
operator of the three sets $B$, $C$ and $D$ is
already the ${}^1 P_1$
combination; the other three may be rewritten as
linear combinations
\begin{equation} \label{onia2tm}
\left(\begin{array}{c} {}^3 P_0 \otimes {}^3 P_0 \\
                       {}^3 P_1 \otimes {}^3 P_1 \\
                       {}^3 P_2 \otimes {}^3 P_2
\end{array} \right)
= \left( \begin{array}{ccc}
  1/3  & 0    & 0   \\
  0    &  1/2 & -1/2   \\
  -1/3 &  1/2 & 1/2 \\ \end{array} \right)
\left(\begin{array}{c} \nabla_\mu \gamma^\mu \otimes
\nabla^\nu \gamma_\nu \\
                       \nabla_\mu \gamma_\nu \otimes
\nabla^\mu \gamma^\nu \\
                       \nabla_\mu \gamma_\nu \otimes
\nabla^\nu \gamma^\mu
\end{array} \right)
\ .
\end{equation}
By inverting (\ref{onia2tm}) we have
\begin{equation}
\left(\begin{array}{c} \nabla_\mu \gamma^\mu \otimes
\nabla^\nu \gamma_\nu \\
                       \nabla_\mu \gamma_\nu \otimes
\nabla^\mu \gamma^\nu \\
                       \nabla_\mu \gamma_\nu \otimes
\nabla^\nu \gamma^\mu
\end{array} \right)
= \left( \begin{array}{ccc}
  3   & 0    & 0    \\
  1   &  1   & 1  \\
  1   &  -1  & 1   \\ \end{array} \right)
\left(\begin{array}{c} {}^3 P_0 \otimes {}^3 P_0 \\
                       {}^3 P_1 \otimes {}^3 P_1 \\
                       {}^3 P_2 \otimes {}^3 P_2
\end{array} \right)
\ .
\end{equation}
Explicitly, for the $D$-operators we have, for example,
\begin{eqnarray}
 D^{(c)}({}^1P_1) & = & D_1^{(c)}
\nonumber\\
 D^{(c)}({}^3P_0) & = & \frac{1}{3}\, D_2^{(c)}
\nonumber\\
 D^{(c)}({}^3P_1) & = & \frac{1}{2}\,
       \left\{ D_3^{(c)} - D_4^{(c)}\right\}
\nonumber\\
 D^{(c)}({}^3P_2) & = & \frac{1}{2}\,
       \left\{ D_3^{(c)} + D_4^{(c)}\right\}
     - \frac{1}{3} \, D_2^{(c)}
\ .
\label{physops}
\end{eqnarray}
Similarly we can define operators with appropriate
$S$-wave quantum numbers
\begin{eqnarray}
 E^{(c)}({}^1S_0) & = & \frac{1}{2}\, E_1^{(c)}
\nonumber\\
 E^{(c)}({}^3S_1) & = & \frac{1}{2}\, E_4^{(c)}
\nonumber\\
 E^{(c)}({}^3S_1,{}^3D_1) & = & \frac{1}{4}\, \left( E_2^{(c)}
       + E_3^{(c)} \right) -  \frac{1}{6}\, E_4^{(c)}
\ .
\label{Swavephys}
\end{eqnarray}

\end{document}